\documentclass[12pt,a4paper]{article}
\usepackage[colorlinks,linkcolor=blue,anchorcolor=blue,citecolor=blue,urlcolor=blue]{hyperref}
\usepackage{graphicx}
\usepackage{amsmath}
\usepackage{longtable}
\begin{document}

\title{$R$ measurement and QCD studies at future super $\tau-c$ factory}

\author{
Guangshun Huang\footnote{E-mail: hgs@ustc.edu.cn}, 
Wenbiao Yan and Xiaorong Zhou\\
(On behalf of STCF working group)\\
University of Science and Technology of China, 
Hefei 230026, China}
\date{}
\maketitle
\vspace{-9cm}
\vbox{\hfill \large for Snowmass RF03}
\vspace{9cm}

\vspace{1cm}

\begin{abstract}
We review status of $R$ measurement and QCD studies at low energy range,
discuss prospects for a super $\tau-$charm factory in $2-7$ GeV.
With a high-luminosity $e^+e^-$ collider, statistics are no longer problem
for $R$ measurement and a precision of $2\%$ or even better is foreseen,
that will lead to bring down the uncertainty of hadronic contribution 
to the QED running coupling constant $\Delta\alpha_{had}$ 
and the anomalous magnetic moment of the muon ($a_{\mu}$); 
measure the strong coupling constant $\alpha_s$ and the charm quark mass; 
improve the measurement of the resonance parameters of heavy charmonia.
Huge data samples in $2-3$ GeV will make it possible to study 
excited states of $\rho$, $\omega$ and $\phi$, or exotic $Y(2175)$;
measure electromagnetic form factor of mesons and baryons;
and measure fragmentation functions of hadrons.
\end{abstract}

\section{Introduction}
Super $\tau-$charm factories (STCF) have been proposed in 
China~\cite{STCF} and Russia~\cite{SCTF}, presumably to
work in $2-7$ GeV, which is a bridge between the perturbative 
and non perturbative energy region. 
It is therefore an important area that is of particular interest 
for testing QCD predictions. 
The STCF will be one of the crucial precision frontier for 
exploring the nature of non-perturbative strong interactions. 
The experimental data will provide essential information to study 
QCD dynamics of confinement through the study of hadron spectroscopy. 
Specifically, high-statistics data will significantly improve 
the following measurements and studies:
\begin{itemize}
\item $R$ Measurement
\item Measurement of the strong running coupling constant $\alpha_s$;
\item Determination of the mass of the charm quark;
\item Production cross section of exclusive hadronic channels and the
electric and magnetic form factors of light baryons;
\item A unique window for charmed baryons;
\item Measurement of the inclusive distribution $x$ and $\xi$;
\item Topological event shapes, such as multiplicity, sphericity and thrust.
\end{itemize}

The Chinese version of STCF is a symmetric electron-positron collider
designed to provide $e^{+}e^{-}$ interactions at $\sqrt{s}=2.0\sim7.0$~GeV. 
The peaking luminosity is expected to be of
$0.5\times10^{35}$~cm$^{-2}$s$^{-1}$ or higher at $\sqrt{s}=4.0$~GeV. 
The proposed STCF would leave space for higher luminosity upgrades 
and for the implementation of a longituinal polarized $e^{-}$ beam 
in a phase-II project.

The STCF detector, a state-of-the-art $4\pi$-solid-angle particle detector
operating at a high luminosity collider, is a general-purpose detector.  
It incorporates a tracking system composed of an inner tracker and main drift
chamber, a particle identification system, an electromagnetic calorimeter, 
a super-conducting solenoid, and a muon detector at the outmost.  To fully
exploit the physics opportunities and cope with the high luminosity, the
STCF is designed with following requirements: (nearly) $4\pi$ solid angle
coverage for both charged and neutral particles, and uniform response for
these final states; excellent momentum and angular resolution for charged
particles, with $\sigma_{p}/p=0.5\%$ at $p=1$~GeV/c; high resolution of
energy and position reconstruction for photons, with
$\sigma_{E}/E\approx2.5\%$ and $\sigma_{\rm pos}\approx 5$~mm at $E=1$~GeV;
superior particle identification ability ($e/\mu/\pi/K/p/\gamma$ and other
neutral particles) and high detection efficiency for low momentum/energy
particles; precision luminosity measurement; tolerance to high
rate/background environment.

\section{$R$ measurement}
According to quark-parton model, hadrons produced via $e^+e^-$ collision are
characterized by the annihilation of $e^+e^-$ into a virtual $\gamma^*$ 
or $Z^0$ boson.  In the lowest order, the cross section for
the (QED) processes $e^+e^- \to \gamma^* \to q\bar{q}$  is related to that for
$e^+e^- \to \gamma^* \to \mu^+ \mu^-$,
\begin{equation}
\sigma(e^+e^- \to q\bar{q}) = 3 \sum_f Q_f^2 \sigma(e^+e^- \to \mu^+ \mu^-)
\end{equation}
where $Q_f$ is the fractional charge of the quark, and three in front records
the three colors for each flavor.  Summing over all the quark flavors, one
defines the ratio of the rate of hadron production to that for muon pairs
as
\begin{equation}
R \equiv \frac{\sigma(e^+e^- \to \gamma^* \to hadrons)}
              {\sigma(e^+e^- \to \gamma^* \to \mu^+ \mu^-)}
= 3 \sum_f Q_f^2.
\label{Rdef}
\end{equation}

The cross section of the pure QED process $e^+e^- \to \gamma^* \to \mu^+ \mu^-$ 
can be precisely calculated, which is the Born cross section 
$\sigma(e^+e^- \to \gamma^* \to \mu^+ \mu^-)=4\pi\alpha^2/3s$. 
Thus, a measurement of total $e^+e^-$ annihilation cross section 
into hadron counts directly the number of quarks, their flavor and colors.  
The $R$ value is expected to be constant so long as the center-of-mass 
energy of the annihilated $e^+e^-$ does not overlap with resonances 
or thresholds for the production of new quark flavors. 

\subsection{Motivation of the precision measurement of the $R$ values}
\subsubsection{$\alpha(M_Z^2)$ and the Standard Model fits}
A remarkable progress has been made in precision test of the Standard Model
(SM) during the past thirty years.  For the analysis of electroweak
data in the SM\cite{EW1, EW2} one starts from the input parameters.  Some of
them, like $\alpha(M_Z^2)$, $G_F$, and $M_Z$ are very well known, 
and some others, $m_{light}$ and $\alpha_s(M_Z^2)$ are only approximately 
determined while $m_t$ and $m_H$ are still poorly known.  
Constrain on $m_t$ and $m_H$ can be derived by comparing the measured 
observables with theoretical predictions that has been calculated to 
full one-loop accuracy and partial two-loop precision, a sufficient 
precision to match the experimental capabilities.
 
Out of the three accurately determined quantities 
$\alpha(M_Z^2)$, $G_F$, and $M_Z$,
the largest uncertainty comes from the running of QED coupling 
constant $\alpha(s)$ from $s = 0$, where it is known to 0.68 ppb, 
up to the $Z$ pole, which is the scale relevant for the 
electroweak precision test.
When relating measurements performed at different energy scales, 
and if the relation involves $\alpha(s)$, one has to know the 
running of $\alpha(s)$ in different energy scale.  
The uncertainty in $\alpha(M_Z^2)$ arises from the contribution of light
quarks to the photon vacuum polarization 
$\Delta\alpha(s)=-\Pi_{\gamma\gamma}(s)$ at the $Z$ mass scale.  
They are independent of any particular initial or final states and
can be absorbed in $\alpha(s)$
\begin{equation}
\alpha(s) \equiv \frac{\alpha}{1+\Pi_{\gamma\gamma}(s)}
\end{equation}
where $\alpha=1/137.035999679(94)$, which is the fine-structure constant, 
at the precision of 0.68 ppb, and 
\begin{equation}
\Delta \alpha(s) \equiv \frac{\alpha(s)-\alpha}{\alpha(s)}
=-\Pi_{\gamma\gamma}(s)
\end{equation}
$\Delta \alpha$ receives the contribution of the leptonic loops and 
the quark loops to the running
\begin{equation}
\Delta \alpha = \Delta \alpha_l + \Delta \alpha_{had}
\end{equation}
where the leptonic part $\Delta \alpha_l$ can be calculated 
analytically and is well known.  
The hadronic part $\Delta \alpha_{had}$ cannot be entirely 
calculated from QCD because of ambiguities in defining the 
light quark masses $m_u$ and $m_d$ as well as the inherent 
non-perturbative nature of the problem at small energy scale.  
An ingenue way\cite{alphaR} is to relate $\Delta \alpha_{had}$ from 
quark loop diagram to $R_{had}$, as defined by Eqn.\ref{Rdef}
\begin{equation}
{\rm Im} \Pi_{\gamma\gamma}(s) = -\frac{\alpha}{3}R_{had}
\end{equation}
\begin{equation}
{\rm Re} \Pi_{\gamma\gamma}(s) = \frac{\alpha s}{3\pi}
P \int_{4m_\pi^2}^{\infty}ds'\frac{R_{had}(s')}{s'(s'-s)},
\label{partReal}
\end{equation}
where $P$ is the principal value of the integral.

\begin{figure}[htbp]
\centering
\includegraphics[width=\textwidth]{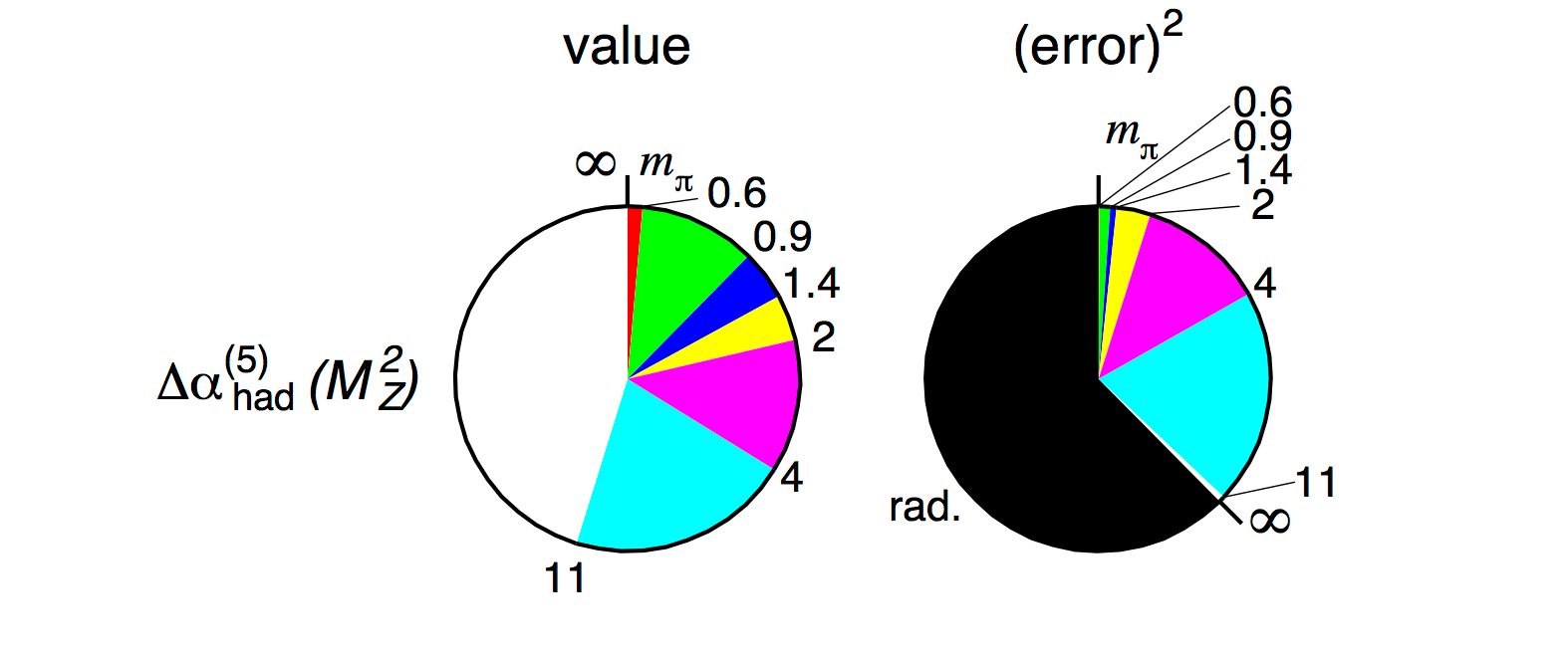}
\caption{Relative contributions to $\Delta \alpha_{had}^{(5)}(M_Z^2)$
in magnitude (left) and uncertainty (right)~\cite{HVP2018}.}
\label{pie_delAlpha_2018}
\end{figure}

Fig.\ref{pie_delAlpha_2018} shows the relative contributions to 
$\Delta \alpha_{had}^{(5)}(M_Z^2)$ in magnitude and uncertainty~\cite{HVP2018}.
The uncertainty of $R$ values in the energy region of $2-5$ GeV 
is still the second largest contribution to the uncertainty of 
$\Delta \alpha_{had}^{(5)}$.  
As great effort has been made for improving the $R$ measurement 
in the energy region of $1-2$ GeV, the measurement of $R$ in $2-5$ GeV 
is becoming more and more important again.

\subsubsection{g-2 of the lepton}
According to the Dirac theory, a lepton is point-like particle and possesses
a magnetic moment
\begin{equation}
\mu = g \mu_B s,
\end{equation}
where $\mu_B = e \hbar / 2m_e c$ is Bohr magneton and $s$ the lepton spin. 
$g = 2$ for particles of $s = 1/2$ is predicted by the Dirac theory.
 
Anomalous magnetic moment of leptons $a_{lepton} \equiv (g-2)/2$ receives
radiative contributions that can in principle be sensitive to new degree of
freedom and interactions.  The weak interaction and the vacuum polarization
effects are too small to observe for electron because of $m_l^2$-dependence. 
The measurement of $a_\tau$ is very difficult due to its short lifetime.  
However, benefited from its larger mass and relatively long lifetime 
the anomalous magnetic moment of muon $a_\mu$ has been measured with 
very high precision at the CERN Muon Storage Ring\cite{mucern1, mucern2, mucern3}, and 
recently by E821 experiment done at Brookhaven National Laboratory 
to a precision of 0.5 ppm\cite{E821}, which is one of the best measured 
quantities in physics.  Theoretically, $a_\mu$ is
sensitive to large energy scales and very high order radiative corrections
\cite{amutheo1, amutheo2}.  
It therefore provides an extremely clean test of electroweak
theory and may give us hints on possible deviations from the SM
\cite{amutest1, amutest2, amutest3}.

According to different source of contribution, $a_\mu$  can be
decomposed as
\begin{equation}
a_{\mu}^{the} = a_{\mu}^{QED}+a_{\mu}^{had}+a_{\mu}^{weak}+a_{\mu}^{new}.
\end{equation}
The QED contribution, the largest term among all, has been calculated 
to ${\cal O}(\alpha^5)$, including the contribution from $\tau$ 
vacuum polarization.  $a_{\mu}^{weak}$ includes the SM effects 
due to virtual W, Z and Higgs particle exchanges. 
$a_{\mu}^{had}$ denotes the virtual hadronic (quark) contribution 
determined by QCD, part of which corresponds to the effects 
representing the contribution of running $\alpha(s)$ from low energy 
to high energy scale.  It cannot be calculated from first principle 
but relates to the experimentally determined $R_{had}(s)$ through 
the expression
\begin{equation}
\alpha_{\mu}^{had} = (\frac{\alpha m_{\mu}}{3\pi})^2
\int_{4m_{\pi}^2}^{\infty}ds\frac{R_{had}(s) K(s)}{s^{2}}
\label{amuhad}
\end{equation}
where $K(s)$ is a kernel varying from 0.63 at $s=4m_\pi^2$ to 
1.0 at $s=\infty$. 
$a_{\mu}^{new}$ stands for the possible contributions beyond the SM, 
which is assumed to be zero so far.

\begin{figure}[htbp]
\centering
\includegraphics[width=\textwidth]{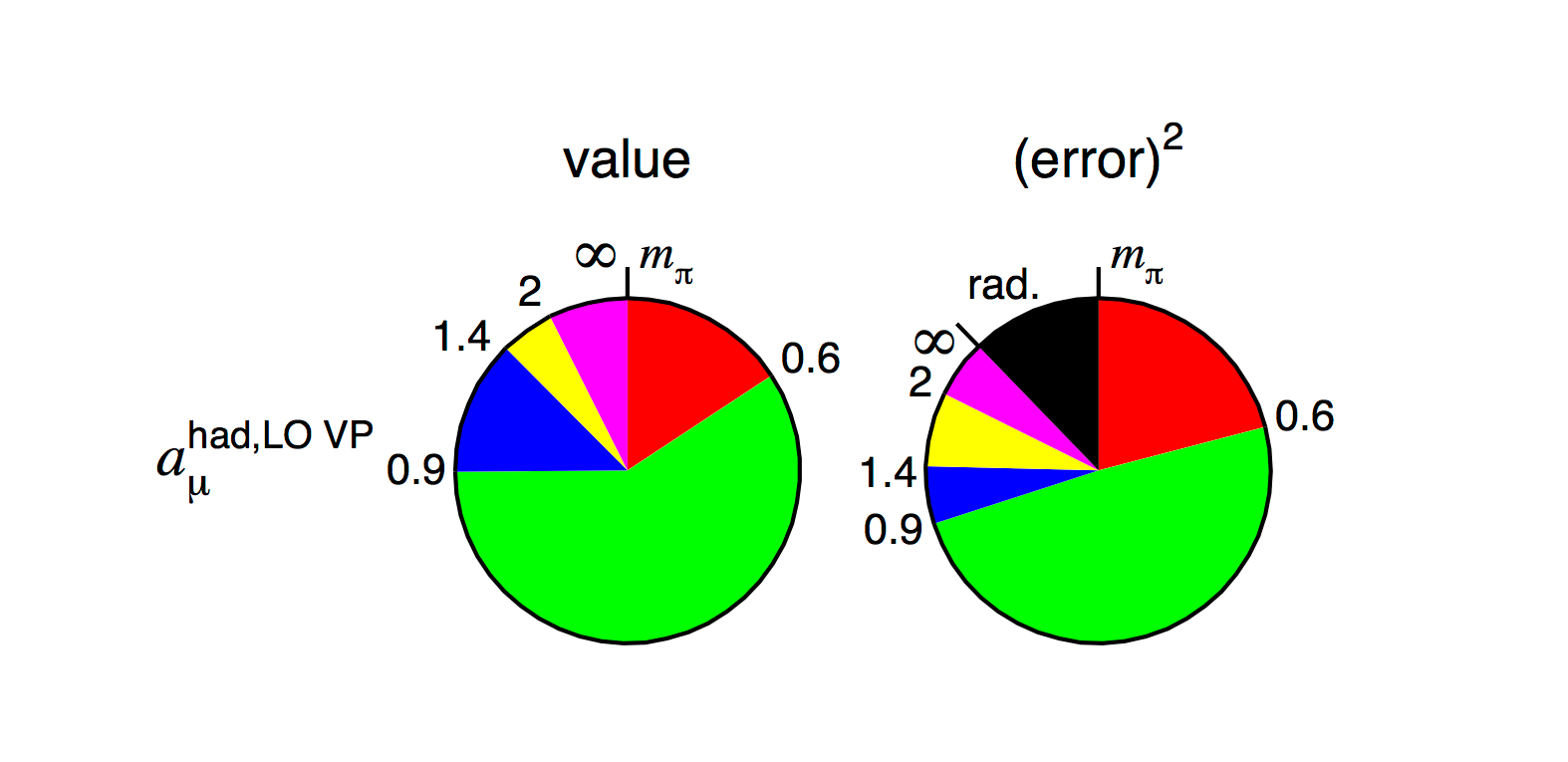}
\caption{Relative contributions to $a_{\mu}^{\rm had}$
in magnitude (left) and uncertainty (right)~\cite{HVP2018}.}
\label{pie_amu_2018}
\end{figure}

The fractional contributions to the total mean value and uncertainty 
of $a_{\mu}^{\rm had}$ from various energy intervals is shown in 
Figure~\ref{pie_amu_2018}~\cite{HVP2018}.

The hadronic vacuum polarization is the most uncertain one of 
all the SM contributions to $a_\mu$.  For several scenarios, it has
been claimed 30 years ago that "the physics achievement of the effort to
re-measure the cross section of $e^+e^-$ hadrons that brings down the
uncertainty of $a_\mu$ to $60 \times 10^{-11}$ is equivalent to that 
of LEP2 or even LHC"\cite{amutest1,amutest4}.  
Now after 30 years effort from both theory and experiments,
the uncertainty on $a_\mu$ has been evaluated to be 
$a_\mu ({\rm SM})= 116 591 810 (43) \times 10^{-11}$~\cite{amuSM2020}, 
bringing down the uncertainty to 0.37 ppm.
The new experiment P989 at Fermilab measured $a_\mu$ to be
$116 592 040 (54) \times 10^{-11}$ (0.46 ppm)~\cite{amu2021FNAL}, 
and it becomes $116 592 061 (41) \times 10^{-11}$ 
(0.35 ppm)~\cite{amu2021FNAL} if combined with previous results.
So the theoretical prediction lags behind, once again calls for 
further reducing the uncertainties of the $R$ values 
in the energy region below 5 GeV.

Near the threshold, as seen from Eqn.\ref{partReal} and Eqn.\ref{amuhad}, 
the integration is proportional to $R_{had}/s^2$, whereas the 
$\Delta\alpha(M_Z)$ integration is proportional to $R_{had}/s$.  
This implies that $a_\mu^{had}$ is more sensitive to the lower
energy than to the higher one.  Measurement in the energy region 
of $0.5-1.5$ GeV from VEPP-2M in Novosibirsk and $\phi$ factory 
at DA$\phi$NE greatly contributed to the interpretation of $g-2$ 
measurement at Brookhaven\cite{E821} and Fermi Lab~\cite{amu2021FNAL}.
However, their contribution to the precision determination of 
$\alpha(M_Z)$ is limited. The $R$ value from BESII at BEPC
in the energy region of $2-5$ GeV made the major contribution 
to evaluate $\alpha(M_Z)$, and also partly contributed to the 
interpretation of g-2. New measurement of $R$ value is highly 
anticipated at a future super $\tau-c$ factory.

\begin{figure}[htbp]
\centering
\includegraphics[height=8cm]{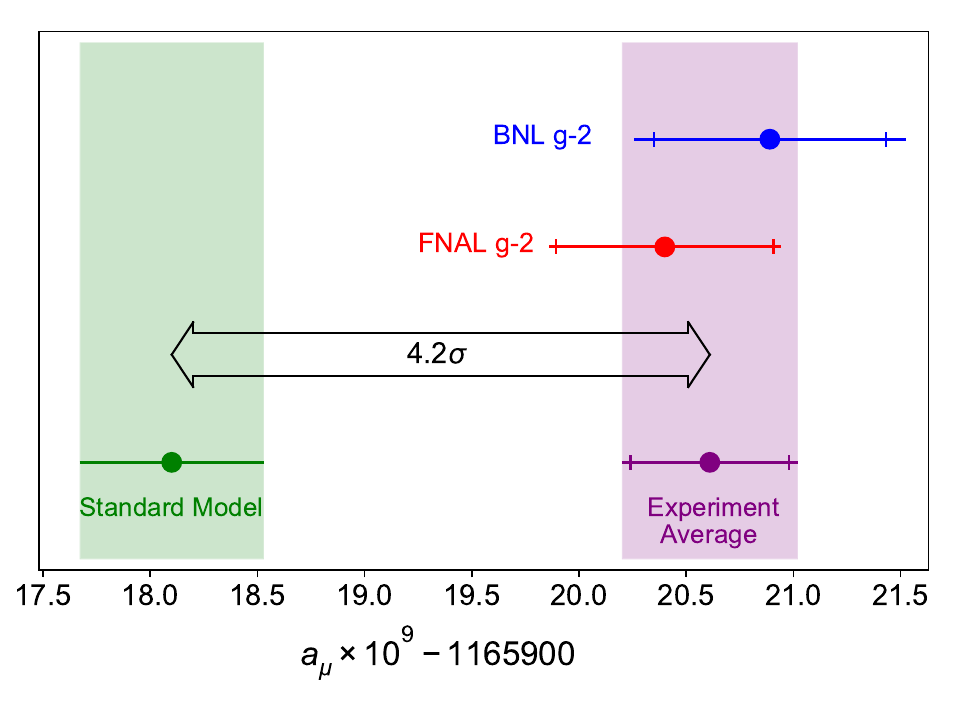}
\caption{Anomalous magnetic moment of leptons
predicted by the SM and measured by the experiment~\cite{amu2021FNAL}.}
\label{amu2021}
\end{figure}

\subsection{Current status and potential to measure $R$}
Experimentally the $R$ value is determined as following, 
\begin{equation}
R = \frac{\sigma_{had}^0}{\sigma_{\mu\mu}^0}
=\frac{ N^{obs}_{had} - N_{bg} - \sum_{l}N_{ll} - N_{\gamma\gamma} }
{\sigma_{\mu\mu}^0 \cdot L \cdot \bar\varepsilon_{had} \cdot 
\varepsilon_{trg} \cdot (1+\delta)},
\label{eqnRexp}
\end{equation}
where $1+\delta$ is a factor taking the radiative correction 
into account for the initial states. $N^{obs}_{had}$ is the 
number of hadronic events collected during a colliding-beam run 
at a certain energy with integrated luminosity $L$ and survived 
after applying the hadronic events selection criteria.  
To obtain $\sigma_{had}^0$, $N^{obs}_{had}$ must be corrected 
for background from different sources.  
The beam associated background $N_{bg}$ will be estimated 
from separated-beam data recorded at each energy to be measured.  
$N_{ll} (l=e, \mu, \tau, \gamma)$, the background contributed 
from lepton pair production and two-photon process can be
estimated by Monte Carlo simulation.

\begin{figure}[htbp]
\centering
\includegraphics[height=8cm,bb=124 285 485 483,clip]{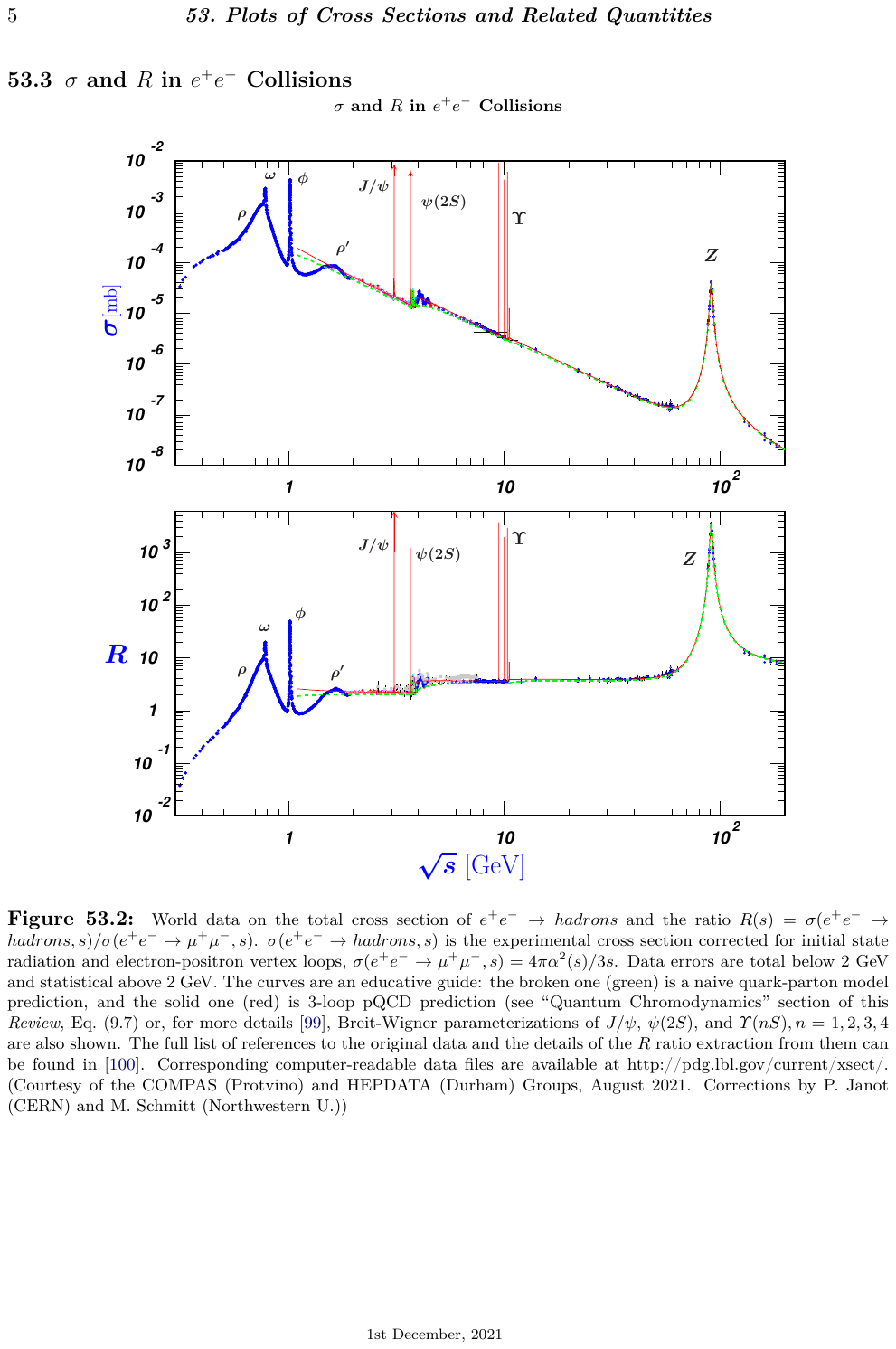}
\caption{R in $e^+e^-$ collision\cite{PDG21}.}
\label{R_PDG2021_all}
\end{figure}

On the other hand, the higher order QCD corrections to $R$ 
has also been calculated in complete 3rd order perturbation 
theory\cite{RQCD}, and the results can be expressed as
\begin{equation}
R=3\sum \limits_{f}Q_f^2[1 + (\frac{\alpha_s(s)}{\pi})
+ 1.411(\frac{\alpha_s(s)}{\pi})^2
- 12.8(\frac{\alpha_s(s)}{\pi})^3+...]
\label{R_alpha_s}
\end{equation}
where $\alpha_s(s)$ is the strong coupling constant. 
Precise measurement of $R$ can be employed to determine $\alpha_s$ 
according to Eqn. \ref{R_alpha_s}, which exhibits a
QCD correction known to ${\cal O}(\alpha_s^3)$.  
In addition, at low c.m. energy, non-perturbative corrections 
(e.g. resonances, etc.) could be important. 
$R$ has been measured by many laboratories in the energy region
covering from hadron production threshold to the $Z^0$ pole\cite{PDG21}.
The results are shown in Fig.\ref{R_PDG2021_all}.  

The uncertainties of $R$ in different energy region are summarized in 
Table \ref{Rstatus}.
The uncertainty of $R$ at the center-of-mass energy between 1 and 2 GeV is
about 10-15\%, for the energy between 2 and 5 GeV, the uncertainty of $R$ has
been reduced from over 15\%\cite{Rref1, Rref2, Rref3, Rref4} to about 6\% 
after a $R$ scan was performed with BESII at BEPC\cite{bes2r}.  
Between the charm and bottom
thresholds, i.e., about $5-10.4$ GeV, $R$ were measured by Mark I, DASP, PLUTO,
Crystal Ball, LENA, CUSB, DESY-Heidelberg, DM-1\cite{DM1}, CLEO\cite{CLEO3}
collaborations. 
Their systematic normalization uncertainties were about $5-10\%$.  
Above bottom threshold, the measurements were from PEP, PETRA and 
LEP with uncertainties of $2-7\%$.
\begin{table}[htbp]
\caption{Typical uncertainties of $R$ in different
energy region\cite{PDG21}.}
\begin{center}
\begin{tabular}{|c|c|c|c|c|c|} \hline \hline
Ecm (GeV) & $<1$ & $1-2$ & $2-5$ & $5-7$ & $10-m_Z$ \\ \hline
$\Delta R/R$ (\%)& 0.9 & 15 & $3\sim6$ & 6 & $2 \sim 7$ \\ \hline
\end{tabular}
\label{Rstatus}
\end{center}
\end{table}

\begin{figure}[htbp]
\centering
\includegraphics[height=6cm,bb=130 406 489 549,clip]{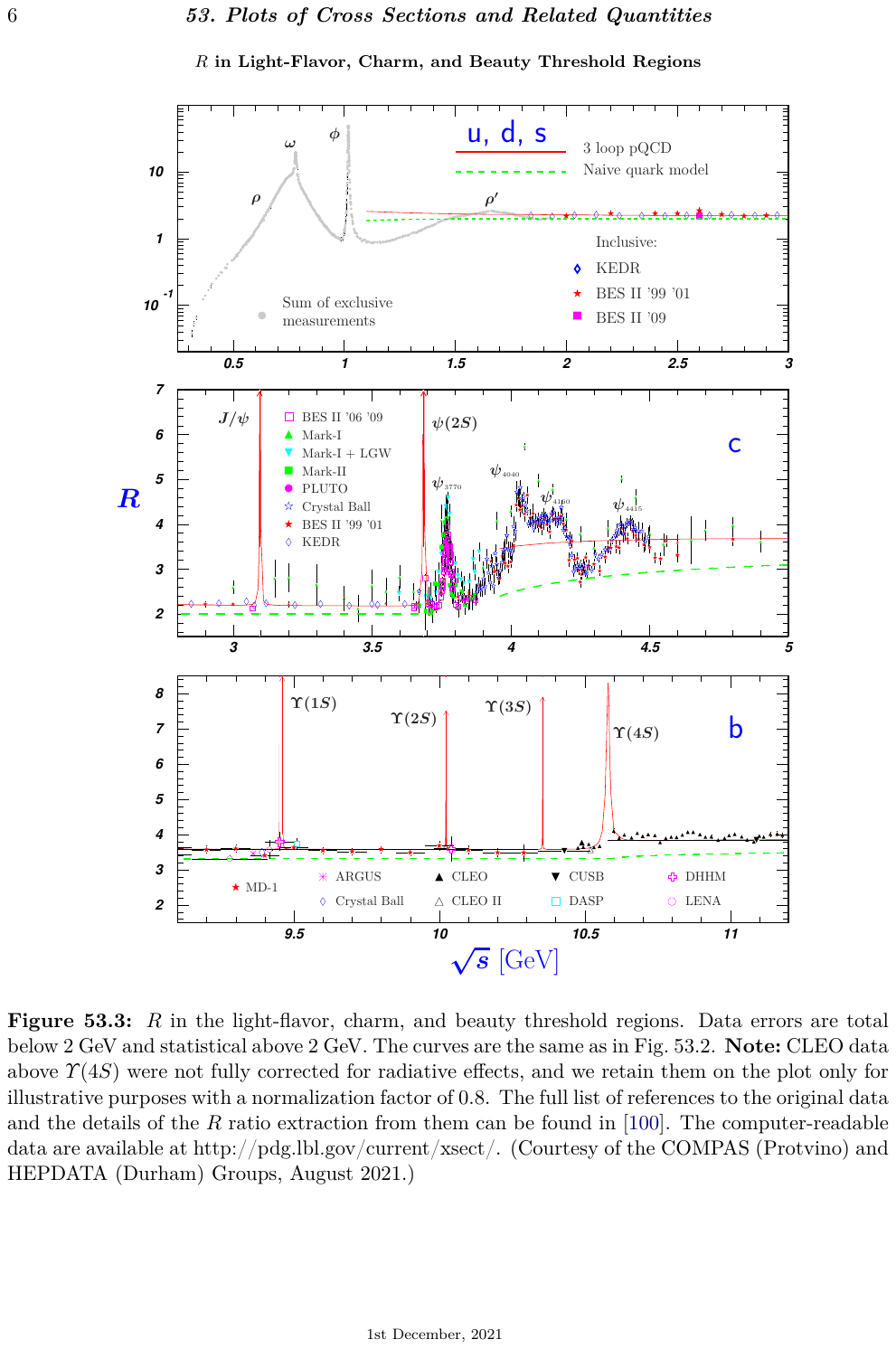}
\caption{R in light-flavor and charm energy region\cite{PDG21}.}
\label{R_PDG2021_low}
\end{figure}

Fig.\ref{R_PDG2021_low} quoted from Ref.\cite{PDG21} shows the $R$ values for
center-of-mass energies up to 5 GeV, including resonances.  The experimental
R values are in general consistent with theoretical predictions, which are
impressive confirmation of the hypothesis of three color degrees of freedom
for quarks.

\begin{figure}[htbp]
\centering
\includegraphics[height=6cm]{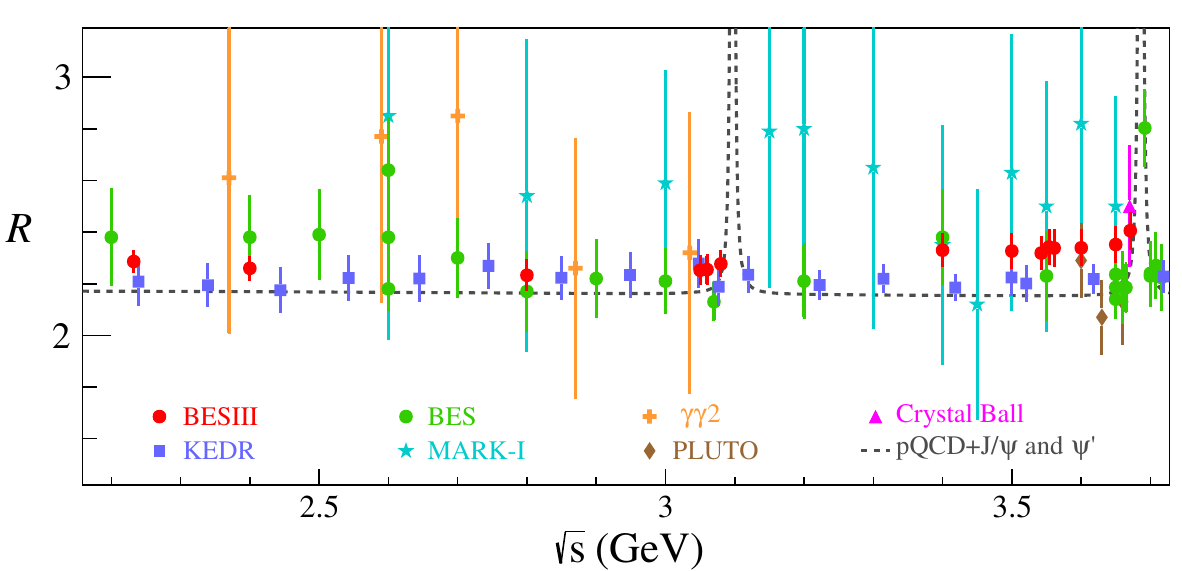}
\caption{R in continuum region\cite{BESIII2022}.}
\label{R2022}
\end{figure}

In recent a few years, KEDR measured $R$ values between 1.84 and 3.72 GeV 
with uncertainties around $3\% \sim 4\%$~\cite{KEDR2016, KEDR2017, KEDR2019}.
BESIII just published its first $R$ measurements between 2.23 and 3.67 GeV
with precision better than $3\%$~\cite{BESIII2022}.
These results are shown in Fig.\ref{R2022}.

\subsection{Heavy charmonia}
\begin{figure}[htbp]
\centering
\includegraphics[height=8cm]{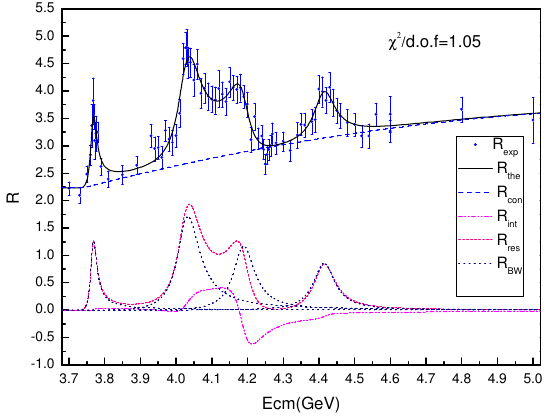}
\caption{Fit to the BESII $R$ spectrum with all two body decays of 
$\psi(3770)$, $\psi(4040)$, $\psi(4160)$ and $\psi(4415)$.
Interference among the resonances and the phase have
significant effect on the fitting.}
\label{fitBES2R}
\end{figure}

DASP group\cite{DASP} inferred the existence of narrow resonance at 4.04 GeV and
4.16 GeV.  In addition to the resonance at 3.77 GeV, Mark I data\cite{MarkI} shown
a broad enhancement at 4.04, 4.2 and 4.4 GeV.  The resonance at 4.4 GeV was
also observed by PLUTO\cite{PLUTO}, but the height and width of the resonance were
reported differently.  The broad resonances at 4.04, 4.16 and 4.45 have been
clearly observed by BES Collaboration from the $R$ scan data, and their
resonance parameters have been measured with improved precision\cite{respara} by
fitting to the inclusive $R$ spectrum as shown in Fig.\ref{fitBES2R}.  
CLEO-c has done a $R$ scan with only 13 energy points, which has shown the
resonances of $\psi(4040)$, $\psi(4160)$\cite{CLEOc13}. 
Using ISR, Belle has reported a measurement of exclusive 
channels and added them up to compare with R\cite{BelleISR}. 

Though the broad resonances at 4040, 4160 and 4415 MeV are assigned to be
the high mass charmonia, we don't have much experimental information for
understanding them.  Their widths are poorly measured, and their decay
channels are not well studied.  Therefore, a high precision $R$ scan in the
$3.9-4.6$ GeV region would be very valuable in disentangling the physics of
that region.

\subsection{R from pQCD and measurement of $\alpha_s$}
A precision measurement of $R$ values is a direct test of QCD. 
According to Eqn.\ref{R_alpha_s}, the strong coupling constant 
$\alpha_s(s)$ can be determined with $R$ values\cite{RQCD, BES09, alphas2020}.

\subsection{Determination of the mass of the charm quark} 
With the help of QCD sum rules, the charm quark mass can be determined
using the $R$ values. There are different ways to define the quark mass.
The one in $\overline{\rm MS}$ scheme is often written as $m_c(m_c)$, or
$m_c(\mu)$ at a chosen energy scale $\mu$ which should be high enough
to ensure convergence of the perturbative series. 
Another form is the pole mass, denoted as $M_c$.
Soon after BESII $R$ values were released, a number of evaluations
of the charm quark mass were reported\cite{cmass1, cmass2}.
The new precision measurement of $R$ values at BESIII will 
once again draw the attention of theorists in the area,
and we may want to determine the charm quark mass by
ourselves as well.

\subsection{Predictions from MLLA/LPHD}
Perturbative QCD can give quantitative
analytical predictions based on the modified leading logarithmic
approximation (MLLA)\cite{MLLA} under the assumption of local 
parton hadron duality (LPHD)\cite{LPHD}.  
At high energies, there are sufficient experimental results, but it
is not the case for low energies.  BESII provided first measurements of
inclusive momenta, multiplicity, the second binomial moments in the energy
region of $2-5$ GeV\cite{yanwb}.  To better test the QCD predictions, 
more accurate measurements with uncertainty of a few percent are 
expected at BESIII.

\subsection{Hadronic form factors}
The electromagnetic form factors (EM FF's) is a fundamental observable of QCD and describes the internal structure of the hadron. It also provides a way to
understand its dynamics.  The proton is one of the basic building blocks of
matter and its EM FF's are necessary for the
interpretation of many theory problems and experimental measurements
involving strong interactions.  The form factors are calculated in the
non-perturbative region in the field theory, where the free parameters in
the semi-phenomenological expression obtained from QCD have to be
measured experimentally.  The proton form factor can be measured in the
space-like region (SL) by studying elastic electron-proton scattering, and the time-like region (TL) by proton-antiproton production in electron-positron annihilation.

The major part of the existing data bank concern SL FF's, while TL FF
measurements are scarce and with large uncertainties. 
BESII measured the proton form factor at 10 energy points in $2-3.07$ GeV, 
but with large statistical uncertainties\cite{lihh}. 
BESIII continued the effort and extracted the form factors of 
proton and neutron with unprecedented precision. 
For more rigorous constraints on QCD phenomenological models,
further improvements are highly desired at future STCF.

For a coherent picture of the low mass spin 1/2 baryon octet, the time-like
form factors of hyperons are also needed: $\Lambda$, $\Sigma$ and $\Xi$. 
BESIII already made contributions, but apparently STCF is more superior with
regard to statistics. Due to the finite life-time of hyperons, it is not
possible to contruct hyperon targets, and observables like space-like form
factors, Transverse Momentum Distributions (TMD's) or Generalized Parton
Distributions (GPD's) are not accessible.  Time-like form factor is
therefore the most powerful way to study hyperon structure.
The key question in hyperon structure
physics is \textit{``What happens, if a light quark in the nucleon is
replaced by a heavier one?''}.  If SU(3) symmetry was exact, the properties
of hyperons could be derived from those of the nucleons.  By comparing the
structure of nucleons and hyperons, we can learn to what extent SU(3)
symmetry is broken and the impact of a heavier quark in a system of light
quarks.  Form factors in the time-like region are complex and the electric
and the magnetic form factor have a relative phase.  This has a polarisation
effect on the final state even if the initial $e^+e^-$ state are
unpolarised.  The weak, parity violating decay of hyperons, that causes the
decay particles to be emmited in the direction of the spin of the hyperon,
makes the polarisation of the hyperon experimentally accessible
\cite{dubnicka}.  This gives a unique feature of the hyperons: the time-like
form factors can be fully determined.

\section{The strangeonium}
The strangeonium ($s\bar{s}$) states~\cite{strangeonium} have not 
been known well like Charmonium ($c\bar{c}$) and bottomonium ($b\bar{b}$).
So far the spectrum of strangeonium has not been established well,
both in theory predictions and in experimental observations.
Below 2.2\,GeV, there should be 22 $s\bar{s}$ resonances expected,
but unfortunately only about half of them are identified.
One reason might be, due to the smaller mass of the $s$ quark,
the strangeonium states are over crowded in the low energy region,
and mixed with hybrids, glueballs and other exotics.

The $\phi(2170)$, most probably a strangeonium state but previously 
referred to as the Y(2175)~\cite{Y2175}, has been 
observed experimentally by {\it BABAR}~\cite{Y2175BABAR11, Y2175BABAR12}, 
Belle~\cite{Y2175BELLE}, BES~\cite{Y2175BESII} and 
BESIII~\cite{Y2175BESIII, Y2175BESIII2019}. 
However the information is diverse, and even the measured mass and width 
of $\phi(2170)$ are controversial. 
There have also been different models for $\phi(2170)$, 
such as traditional $3\ ^3S_1$~\cite{Y2175SSbar2} or 
$2\ ^3D_1$ $s \bar{s}$~\cite{Y2175ss2D} state, 
$1^{--}$ $s\bar{s}g$ hybrid~\cite{Y2175hybrid, Y2175hybrid3}, 
tetraquark state~\cite{Y2175tetraquark1,Y2175tetraquark2, 
Y2175tetraquark3, Y2175tetraquark4, Y2175tetraquark5}, 
$\Lambda\bar{\Lambda}(^3S_1)$ bound state~\cite{Y2175lambda,
Y2175lambda1, Y2175lambda2, Y2175lambda3}, 
$S$-wave threshold effect~\cite{SWaveThreshold},
or $\phi K\bar{K}$ resonance state~\cite{X2170}.

\begin{figure}[htbp]
\centerline{\includegraphics[width=0.6\textwidth]{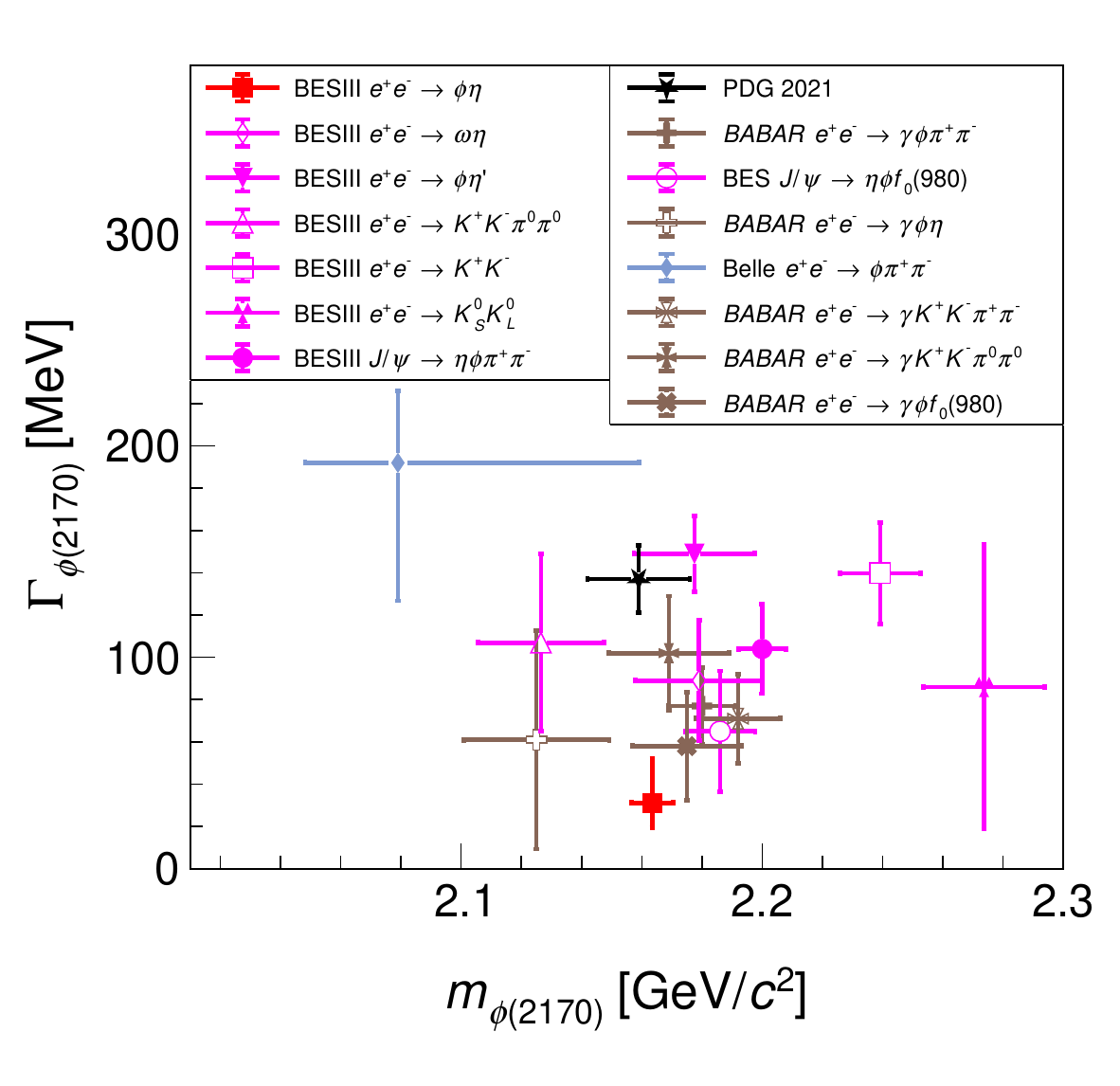}}
\caption{A compilation of measured mass and width of the $\phi(2170)$.}
\label{phi2170para}
\end{figure}

To improve the knowledge, BESIII measured a nmuber of processes,
including $e^+e^-\to K^+K^-$~\cite{KK}, 
$e^+e^-\to K^0_S K^0_L$~\cite{KSKL},
$e^+e^- \to K^+K^-\pi^0\pi^0$~\cite{KKpi0pi0}, 
$e^+e^-\to K^+K^-K^+K^-$/ $\phi K^+K^-$~\cite{KKKK},
$e^+ e^- \to \phi\eta$~\cite{phieta},
$e^+e^-\to \phi\eta^\prime$~\cite{phietap},
$e^+ e^- \to \omega\eta/\pi^0$~\cite{omegaeta},
using data collected in the center-of-mass energy region of $2.0-3.08$ GeV,.
These experimental results provide additional information
in understanding the $\phi(2170)$.
The current situation of the $\phi(2170)$ parameters is displayed
in Fig.~\ref{phi2170para}, indicating the $\phi(2170)$ remains 
intriguing and therefore more efforts are needed. 
Hopefully a STCF will eventually settle it down 
and identify more strangeonium states in the future.

\section{Fragmentation function}
Fragmentation function(FF) $D_q^{h}(z)$ is the probability that
hadron $h$ is found in the debris of a quark (or antiquark) carrying
a fraction $z$ of its energy. The corresponding differential cross 
section can be written as
\begin{equation}
\frac{d\sigma(e^+e^- \rightarrow h + X)}{dz} = 
\sum_{q}\sigma(e^+e^- \rightarrow q \bar{q})(D_q^{h}(z)+D_{\bar{q}}^{h}(z))
\end{equation}
at leading level\cite{martin}. The fragmentation function
$D_q^{h}(z)$ is a nonperturbative object due to hadronization, and
can not be deduced from first principles, but could be extracted
from experimental data on inclusive hadron production.

At BESIII fragmentation functions have being measured 
from inclusive $\pi^0$ and $K_S^0$ productions in the
c.m. energy range between 2.2324 to 3.6710 GeV with
statistical uncertainties comparable to systematic ones.
These measurements fill the area below 10 GeV, 
where there were almost no experimental data.
The results offer unique opportunity to extract the 
unpolarized fragmentation functions in the relatively 
low energy region and to study the QCD dynamics 
in this particular region.

\section{Nucleon form factors}
The ratio of proton electromagnetic form factors ($R_{\rm{em}}$) in the
time-like region (TL) was in poor accuracy ranging between
12\% and 28\%, increasing with increasing momentum transferred by the
virtual photon, $q$.  Before BESIII, there were only two experiments 
which measured $R_{\rm{em}}$: BaBar~\cite{ppbarBaBar1, ppbarBaBar2} 
and PS170~\cite{lear}. 
BaBar measured $R_{\rm{em}}$ in six different $q$-bins via the process
$e^+e^- \rightarrow p\overline{p} \gamma$.  The $q$-region covered by BaBar
was between 1.877 and 3.0 GeV.  PS170 measured the process $p\overline{p}
\rightarrow e^+e^-$ in five fine $q$-bins between 1.931 and 2.049 GeV. 
While the spectrum of the PS170 experiment seems to be compatible with the
assumption $|G_E|/|G_M| = 1$, the BaBar spectrum shows a relatively large
deviation from 1, measuring values of $|G_E|/|G_M|$ greater than unity. 
In the case of the neutrons, there was no independent measurement of 
the electromagnetic form factors or their ratio so far.  
Only FENICE-experiment~\cite{Feni} in Frascati extracted 
the neutron form factor $G^{n}_M$ from the measurement of the
cross section in the reaction $e^+e^-\rightarrow n\overline{n}$ 
and assuming $G^{n}_E = 0$.

\subsection{Nucleon electromagnetic form factors in the TL region}
Electromagnetic form factors (FFs) account for 
the non point-like structure of hadrons. 
The vertex operator $\Gamma^{\mu}(q)$ describing 
the hadronic current in the Feynman diagrams of 
Fig.~\ref{Feynman} can be written in terms of 
the so called Dirac and Pauli FFs, $F_1$ and $F_2$:
\begin{figure}[!t]
\begin{center}
\includegraphics*[width=85mm]{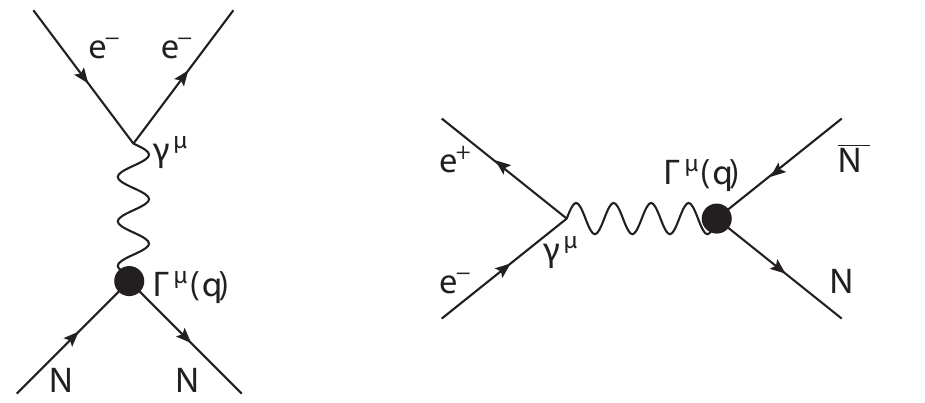}
\end{center}
\caption{\label{Feynman}Lowest-order Feynman diagrams 
for \mbox{$e^- N  \rightarrow e^- N$} (left) and 
for  \mbox{$e^-e^+ \rightarrow N \overline{N}$} (right).}
\end{figure}
\begin{equation}
\Gamma^\mu(q^2) = \gamma^\mu F_1(q^2) + \frac{i\sigma^{\mu\nu}q_\nu}{2m_N} F_2(q^2),
\end{equation}
with $m_N$ the mass of the nucleon $N$ or spin-1/2 baryon. The FFs are analytic functions of the momentum transferred $q^2$. They are real in the space-like (SL) region ($q^2<0$) and can be complex in the time-like (TL) region ($q^2>0$) for $q^2 > 4m_\pi^2$. The use of the so-called Sachs FFs has become conventional: 
\begin{equation}
G_E (q^2)= F_1(q^2) + \frac{q^2}{4m^2}F_2(q^2), 
\hspace{1.25cm}
G_M(q^2) = F_1(q^2) + F_2(q^2),
\end{equation}
with $G_E(0) = G_M(0)/\mu_N$ = 1 and $\mu_N$ the nucleon magnetic moment. Form factors for $q^2<0$ are determined by elastic scattering of electrons from hadrons available as targets. Form factors for $q^2>0$ are measured in  annihilation  processes \mbox{$e^+e^- \leftrightarrow N \overline{N}$}. 
The differential cross section of the annihilation process \mbox{$e^+e^- \rightarrow N\overline{N}$} in c.m.~\cite{crosssection} reads 
\begin{equation}
\label{diff_xs}
\frac{d\sigma(q^2,\theta^{CM}_{N})}{d\Omega} = \frac{\alpha^2\beta C}{4q^2} \left [(1 + \mathrm{cos}^2\theta^{CM}_{N}) |G_M (q^2)|^2 +  \frac{1}{\tau}\mathrm{sin}^2\theta^{CM}_{N}|G_E(q^2)|^2 \right ],
\end{equation}
where $q^2 = M_{p\overline{p}}^2$ is the momentum transferred by the virtual
photon, $\theta^{CM}_{N}$ is the polar angle of the nucleon, $\tau =
4m_{N}^2/q^2$ , $m_N$ is the mass of the nucleon, $\beta = \sqrt{1-1/\tau}$,
$C = y/(1-\mathrm{exp}(-y))$ and $y = \pi \alpha / \beta$.  The Coulomb
factor $C$ accounts for the electromagnetic $N \overline{N}$
interactions~\cite{Tzara}.  The analysis of the angular differential cross
section allows the independent extraction of the electromagnetic for factors
$G_E$ and $G_M$.  Angular integration of the previous equation gives the
total cross section:
\begin{equation}
\sigma(q^2) = \frac{4\pi\alpha^2\beta C}{3q^2} \left[ |G_M(q^2)|^2 + \frac{1}{2\tau} |G_E(q^2)|^2 \right ].
\label{totalcs}
\end{equation}

\subsection{Extraction of $R_{\rm{em}} = |G_E|/|G_M|$ and $G_{E,M}$}
In order to extract the ratio of the electromagnetic form factors, 
the proton angular distributions of the selected events are weighted 
with the selection efficiencies with rad. corr. and fitted with 
the following formula:
\begin{equation} \label{form}
f(\rm{cos}\theta^{CM}_{p}) = \it{Norm} \cdot [\tau (1 + \rm{cos}^2\theta^{CM}_{p}) + \it{R_{\rm{em}}} (1 - \rm{cos}^2\theta^{CM}_{p})], 
\end{equation}
with two free parameters, a global normalization factor, $Norm$, 
and the ratio of the electromagnetic form factors $R_{\rm{em}}$. 
This formula is equivalent to Eq.~\ref{diff_xs} after factorizing out $|G_M|$.
The electromanetic form factors can be extracted as:
\begin{equation}
|G_M|^2 = \frac{Norm\cdot 2 q^2\tau}{\rm{L}\cdot \rm{bw}\cdot\pi\alpha^2\beta},
\end{equation}
\begin{equation}
|G_E|^2 = |R_{\mathrm{em}}|^2 \cdot |G_M|^2, 
\end{equation}
where L is luminosity and bw is bin width. Due to the factor $1/\tau$ 
in front of $|G_E|^2$ in Eq.~\ref{diff_xs}, this form factor is strongly 
suppressed at high $q^2$.

\subsection{ISR vs energy-scan}
\begin{figure}[!t]  
\begin{center}
  \includegraphics[width=16pc]{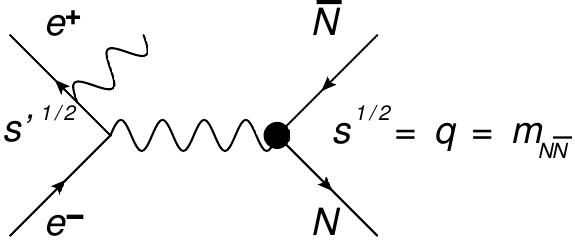}
\caption{One of the lowest-order Feynman diagrams for 
initial-state radiation emission in $e^+e^- \rightarrow p\bar{p}$.}
\label{fig:ISR}
\end{center}
\end{figure}

Taking advantage of the ISR-technique (Fig.~\ref{fig:ISR}), 
the whole range below a fixed nominal energy would be accessible.
As indicated by Eq.~\ref{LISR}, $L_{ISR}(\sqrt{s},E_{\gamma})$, 
and $L_0(\sqrt{s})$ are the ISR-Luminosity at $q = \sqrt{s}$ and 
the luminosity at at the same $q$ for the direct annihilation, respectively. 
The use of the ISR-technique to measure electromagnetic nucleon 
form factors would allow to measure the electromagnetic form factors 
in bins of non-negligible finite size of $q^2$, but not in 
differential $q^2$-bins. 

\begin{equation}
L_{ISR} (\sqrt{s},E_{\gamma})= \beta \frac{dE_\gamma}{E_\gamma} \left [1 - \frac{dE_\gamma}{E_\gamma} + \frac{1}{2} \left (\frac{dE_\gamma}{E_\gamma} \right)^2 \right ] \cdot d\sqrt{s'} \cdot L_0(\sqrt{s}),
\label{LISR}
\end{equation}
with $E_{\gamma} = \sqrt{s'}/2 \cdot (1 - m_{N\bar{N}}^2)$ the energy of the real photon and $m_{N\bar{N}}$ the invariant-mass of the $N\bar{N}$-system.

\subsection{Nucleon form factors at BESIII}
\subsubsection{Proton}
\begin{figure}[tp]
 \centering
 \includegraphics[height=5.5cm]{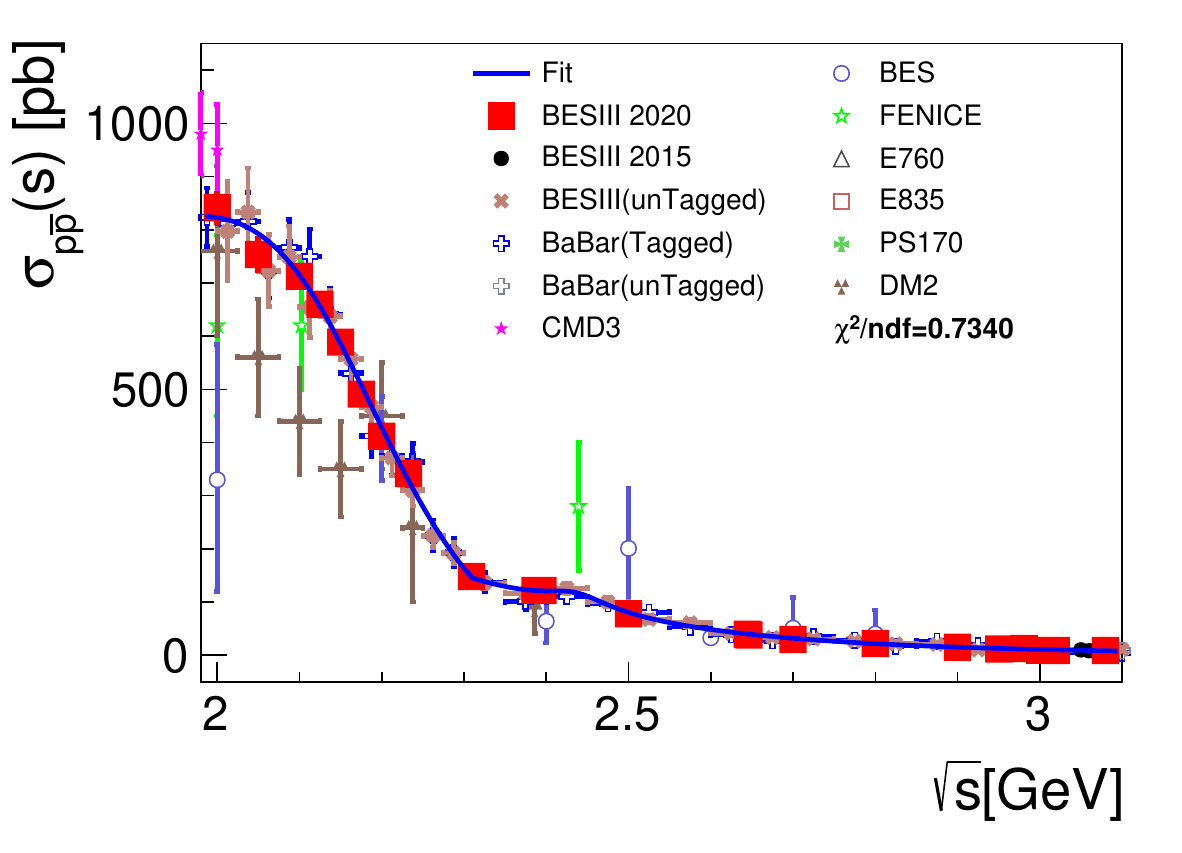}\hspace{0.5cm}%
 \includegraphics[height=5.5cm]{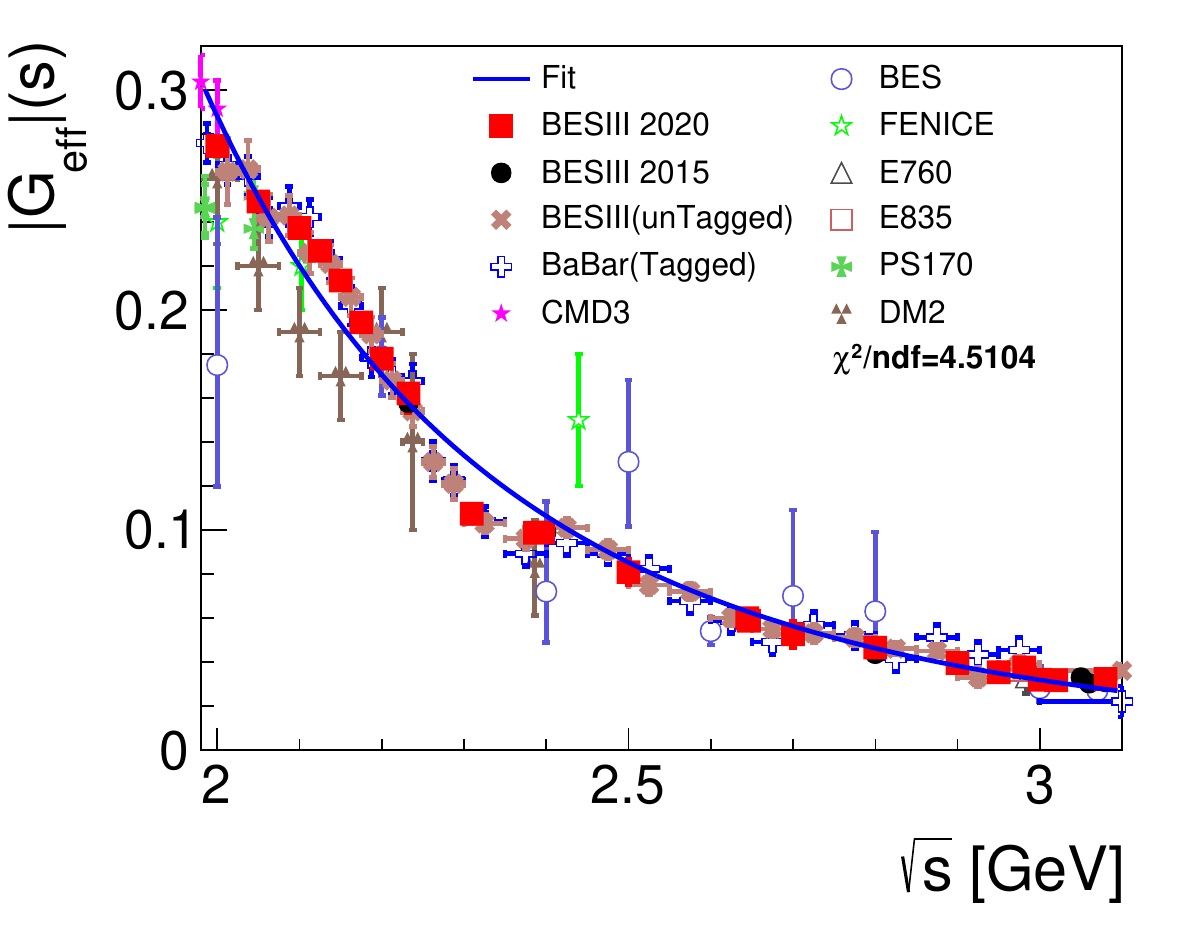}\hspace{0.5cm}%
 \put(-375,135){(a)}\put(-150,135){(b)}\\
 \includegraphics[height=5.5cm]{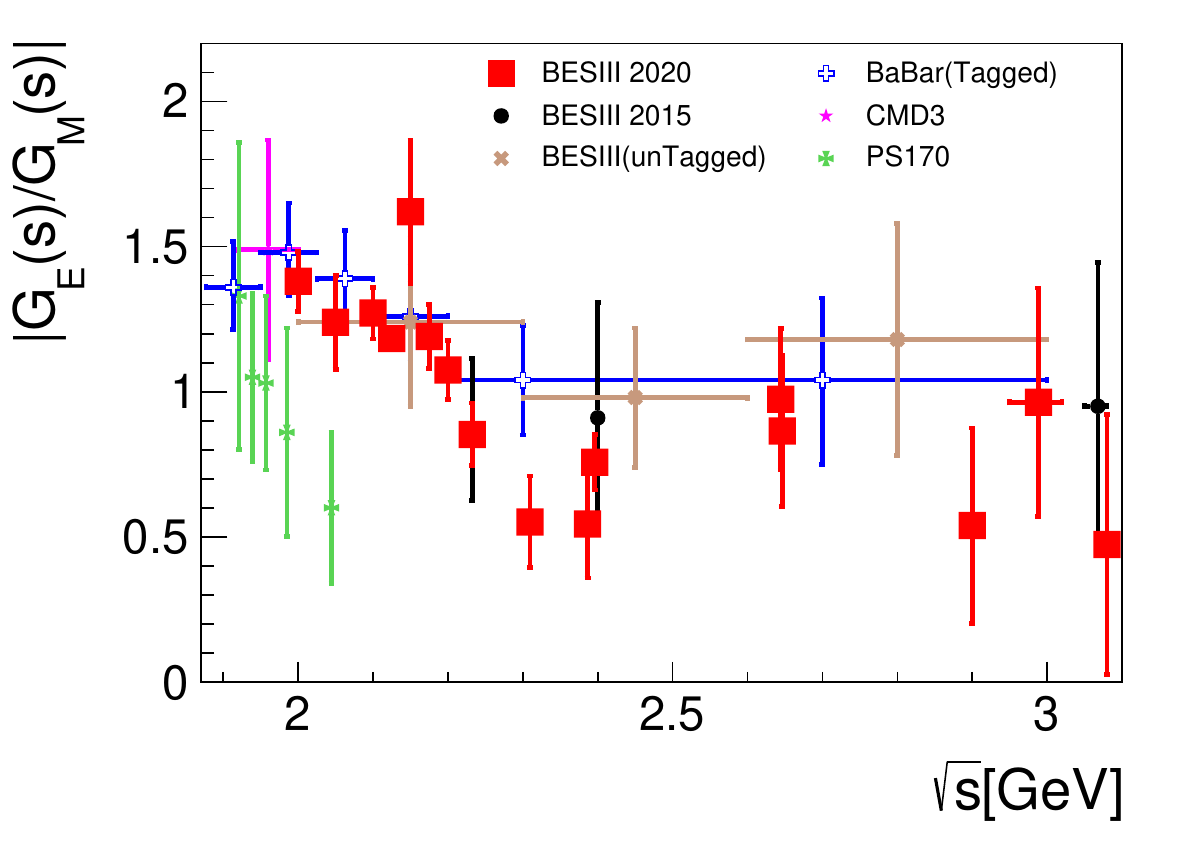}\hspace{0.5cm}%
 \includegraphics[height=5.5cm]{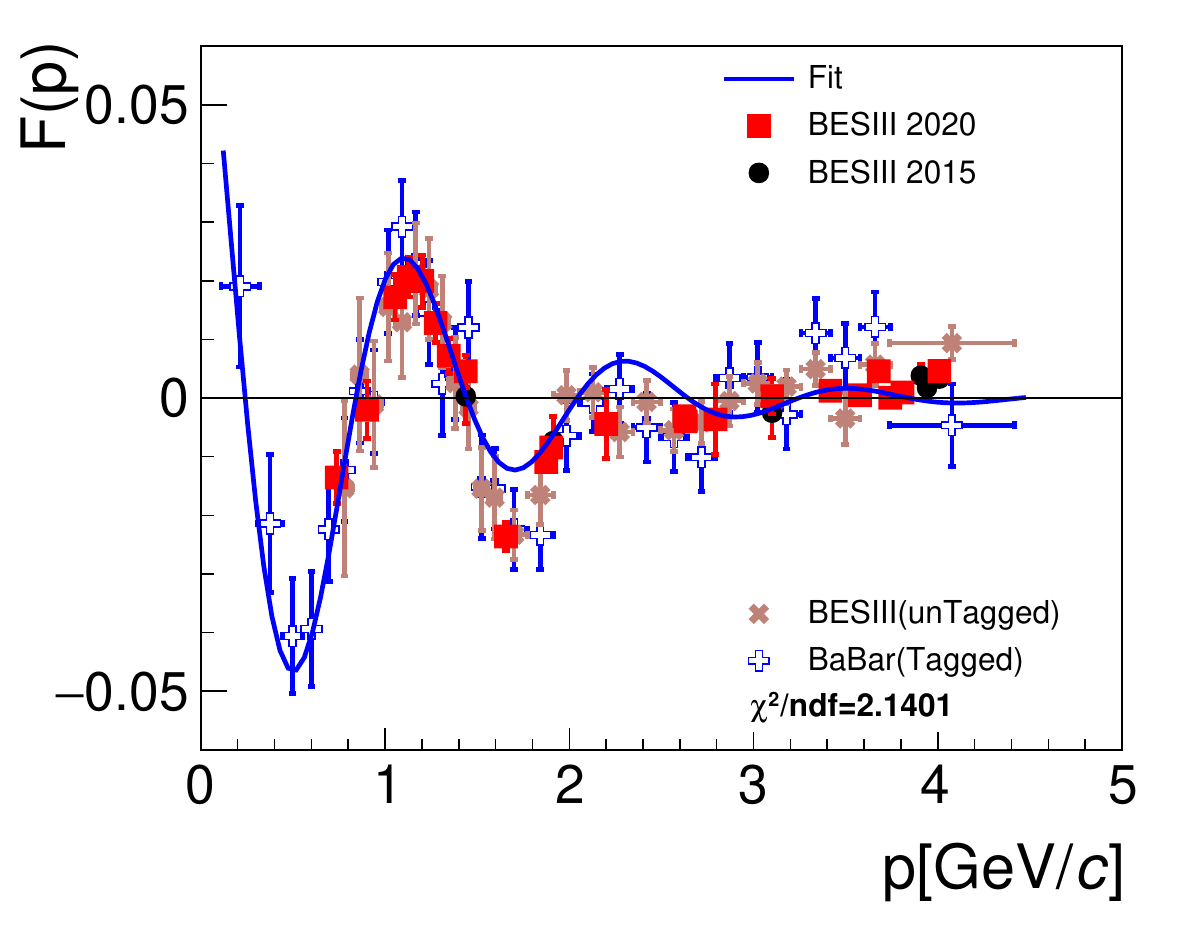}
 \put(-375,135){(c)}\put(-150,135){(d)}
 \caption{\label{fig:proton}
(a) the cross sections for $e^+e^- \to p\bar{p}$ cross section. 
(b) the effective proton time-like form factor. 
The blue curve is the results of an attempt to fit the measurements 
with smooth dipole-like function.
(c) the ratio $R=|G_E/G_M|$. 
(d) effective form factor residual $F(p)$ after subtracting 
the one calculated by QCD theory (the blue curve shown in (b)), 
as a function of the relative motion $p$ of the final proton and antiproton.}
\end{figure}

BESIII obtained the most accurate proton $|G_E/G_M|$ ratio measurements 
at 16 c.m. energies between 2.0 and 3.08 GeV~\cite{ppbar2015, ppbar2020} 
that favor BaBar over PS170 and helped clarifying the puzzle. 
BESIII also performed the measurements using the ISR 
technique~\cite{ppbaruntag, ppbartag}, 
with results that are consistent with BaBar's.
The BESIII measurements are shown in Fig.~\ref{fig:proton} 
(a) for $p\bar{p}$ production cross section in $2.0 - 3.08$ GeV,
(b) the effective proton time-like form factor,
(c) the form factor ratio $R=|G_E/G_M|$, and
(d) the effective form factor residual,
together with results from other experiments.
The electric form factor was extracted for the first time.
The unprecedented 3.5\% uncertainty that was achieved at 2.125 GeV 
by BESIII is close to that of the best measurements in the space-like 
region, which have been at per cent level since long time ago.
The CMD-3 experiment measured the production cross section of proton pair 
and observed an abrupt rise at the nucleon-antinucleon threshold~\cite{CMD3},
as expected for point-like charged particles according to Eqn.~\ref{diff_xs}. 
BESIII did not extend down to the threshold energy, but the results 
around 2 GeV agree with CMD-3.
This information improves our understanding of the proton inner structure 
from a different dimension and helps to test theoretical models 
that depend on non-perturbative QCD, e.g. charge distribution 
within the proton can be deduced~\cite{unitary1, unitary2}.
The near threshold behavior of the electromagnetic form factor 
of a hadron is mostly determined by the interaction of the 
hadron-antihadron in the final state, and therefore the measurements 
of the form factor properties can also serve as a fruitful source of 
information about hadron-antihadron interaction~\cite{Dalkarov}.

Interestingly there are oscillations in the effective proton form factor, 
first seen by BaBar and later confirmed by BESIII~\cite{ppbaruntag}.
These oscillations were subsequently studied with more precise data
by BESIII~\cite{ppbar2020}.
Ref.~\cite{PeriodicFF} speculated that possible origins of this 
curious behavior are rescattering processes at relative distances 
of 0.7 - 1.5 fm between the centers of the forming hadrons, leading 
to a large fraction of inelastic processes in $p-\bar{p}$ interactions, 
and a large imaginary component to the rescattering processes.

\subsubsection{Neutron}
\begin{figure}[htbp]
\includegraphics[width=0.48\textwidth]{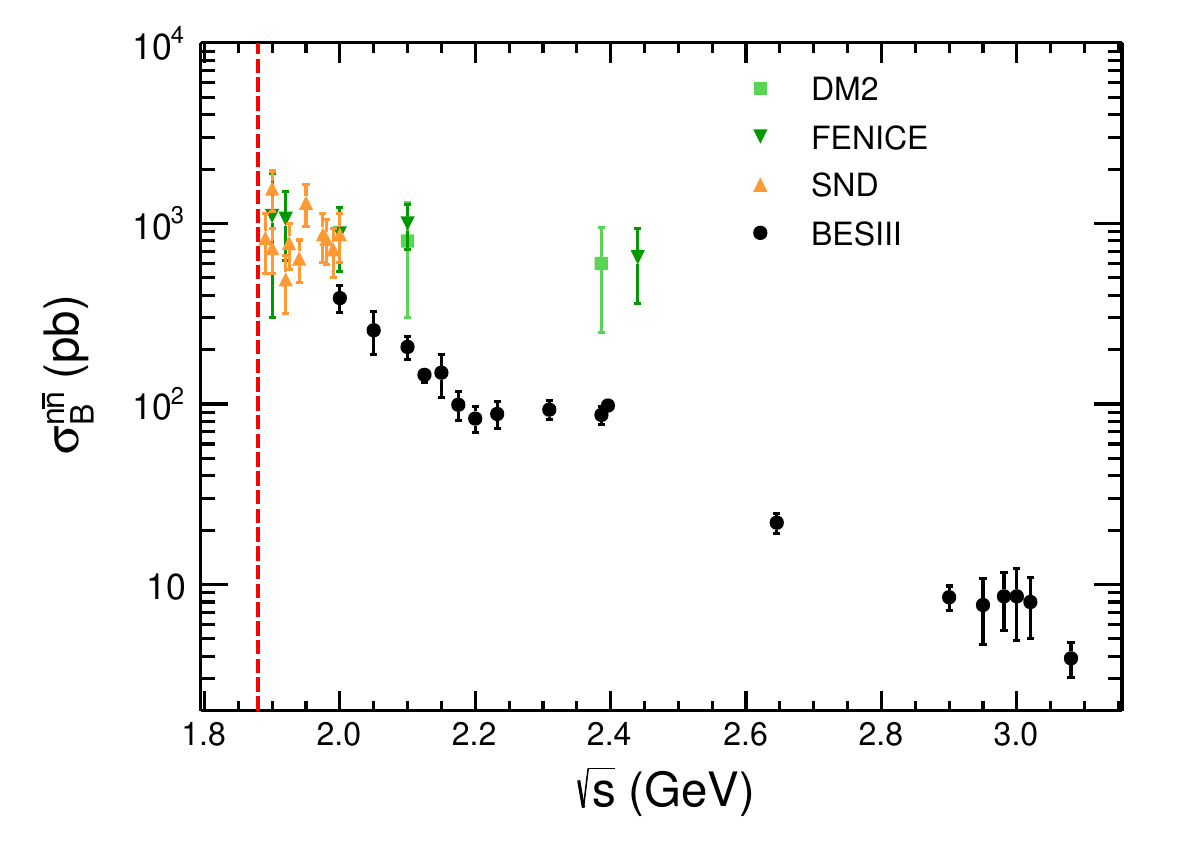}
\includegraphics[width=0.48\textwidth]{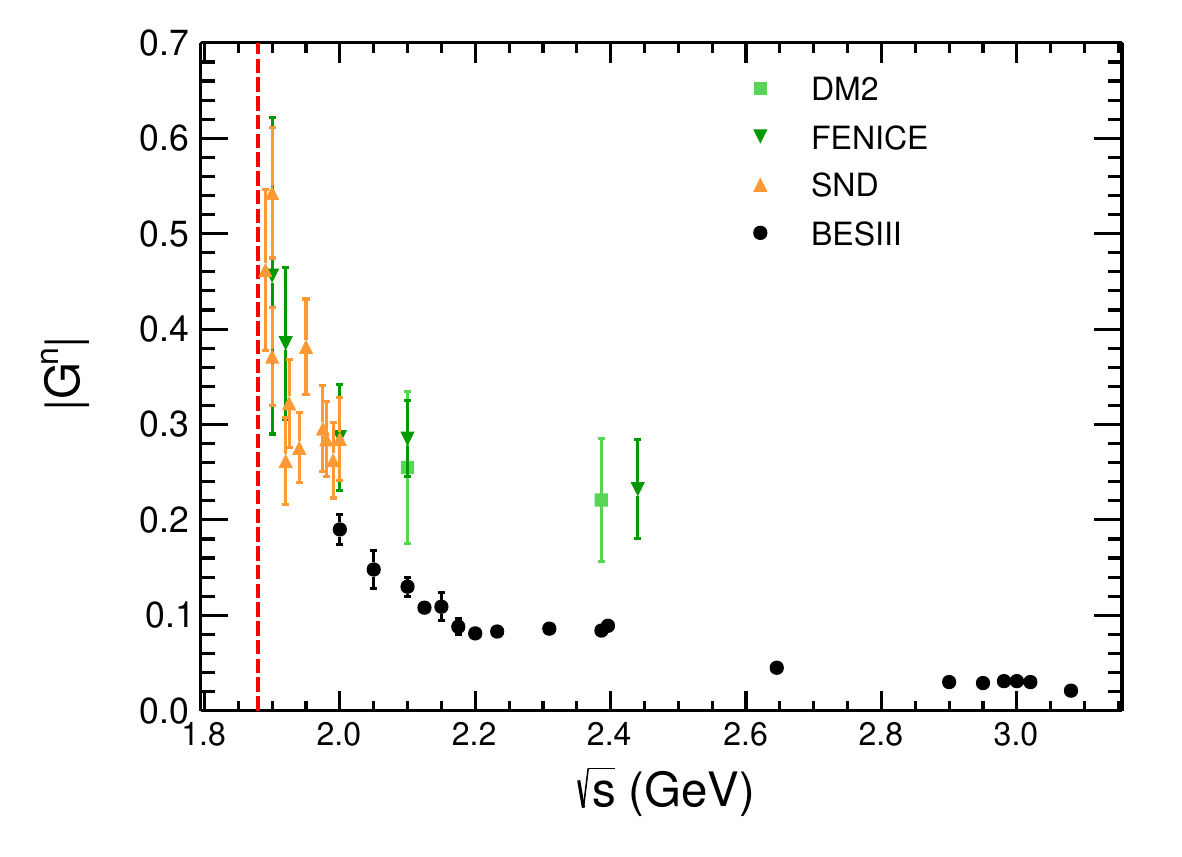}
\caption{\textbf{Left:} Results for the Born cross section 
$\sigma_{B}^{n\bar{n}}$ with respect to the center-of-mass energy $\sqrt{s}$. 
\textbf{Right:} Results for the effective form factor $|G^n|$ 
with respect to the center-of-mass energy $\sqrt{s}$.}
\label{nnGeffXS}
\end{figure}

The most recent measurement of the TL neutron FFs was performed
at the BESIII experiment.
A data set with a total integrated luminosity of 647.9 pb$^{-1}$ 
at 18 energies between $\sqrt{s} = 2.0$ and 3.08 GeV 
was used and over 2000 $n\bar{n}$ events were selected 
to determine $\sigma_B^{n\bar{n}}$ and $|G^n|$. 
Because the final state neutron and anti-neutron
are both neutral, with no tracks recorded in
the drift chamber, the event selection is a challenge.
The information in the calorimeter and the 
time of flight counters has to be used to identify the signal;
as such the selection efficiency is much lower and  
the number of observed neutron events is significantly 
less than that for protons. 
The results were published in 2021~\cite{nnbar} and represent 
the most precise and extensive measurement up to date.
The precision of $\sigma_B^{n\bar{n}}$ is greatly improved 
when compared to previous measurements. 
The accuracy of $\sigma_B^{n\bar{n}}$ ranges between $\sim$4--40\% 
and  $\sim$6--16\% from statistics and systematic effects, respectively. 
The results from BESIII on $\sigma_B^{n\bar{n}}$ and $|G^n|$ are shown 
in Fig.~\ref{nnGeffXS}.

\begin{figure}[htbp]
\begin{center}
\includegraphics[width=0.6\textwidth]{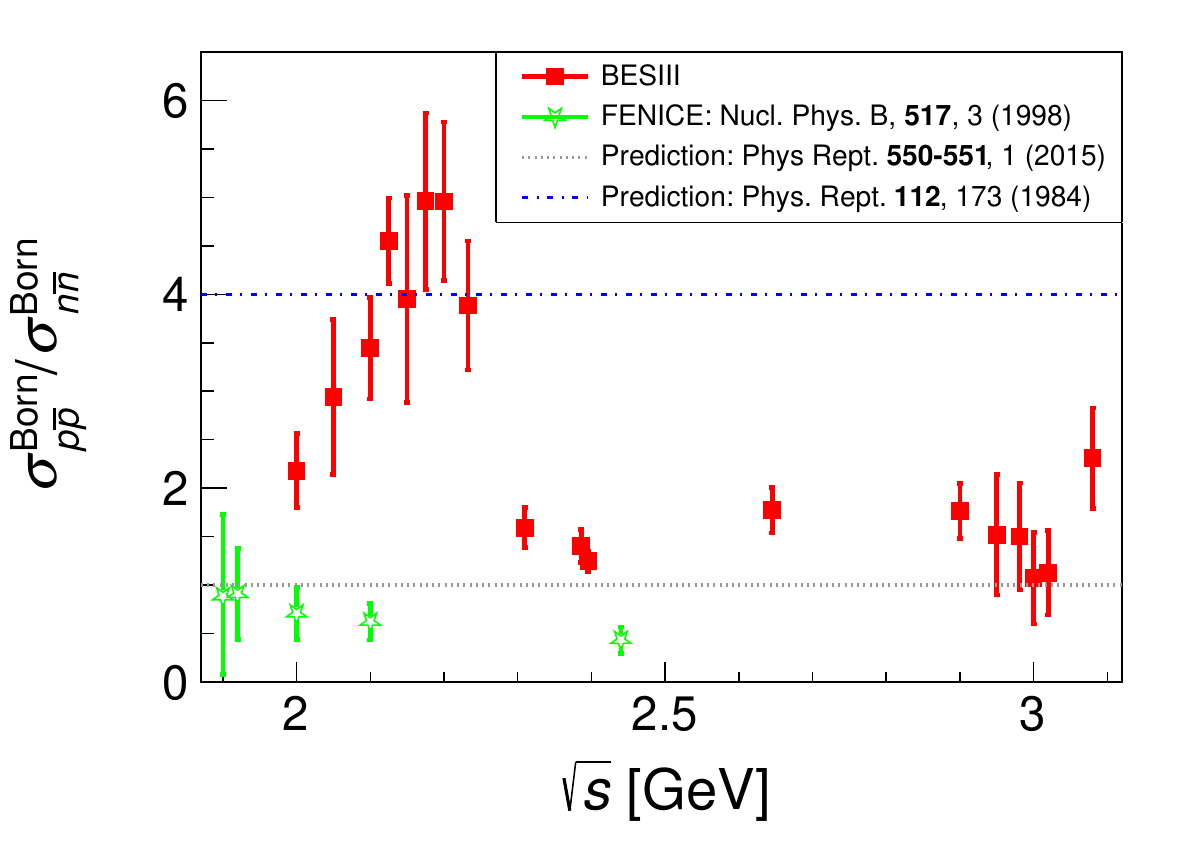}
\caption{Ratio of Born cross section of $e^+e^- \to p\bar{p}$
to that of $e^+e^- \to n\bar{n}$.}
\label{pp2nn}
\end{center}
\end{figure}

Neutron measurements from SND~\cite{SND2014, SNDnn} and BESIII~\cite{nnbar}
overlap and roughly agree at 2 GeV, where a cross-section behavior 
that is close to the $e^+e^- \to p\bar{p}$ case is observed,
in particular a flat behavior above threshold up to 2 GeV 
as seen by CMD-3~\cite{CMD3}, but this challenges 
the expected behavior from Eqn.~\ref{diff_xs}.
For energies above 2 GeV, the BESIII measurements of the
ratio of the proton to neutron cross sections is 
more compatible with the QCD-motivated model predictions:
as shown in Fig.~\ref{pp2nn}, the cross section for $e^+e^- \to p\bar{p}$ 
is larger than for $e^+e^- \to n\bar{n}$ in general.

\section{Hyperon threshold effects}
\label{sec:threshold}
\begin{figure}[htbp!]
\centerline{\includegraphics[width=0.6\linewidth]{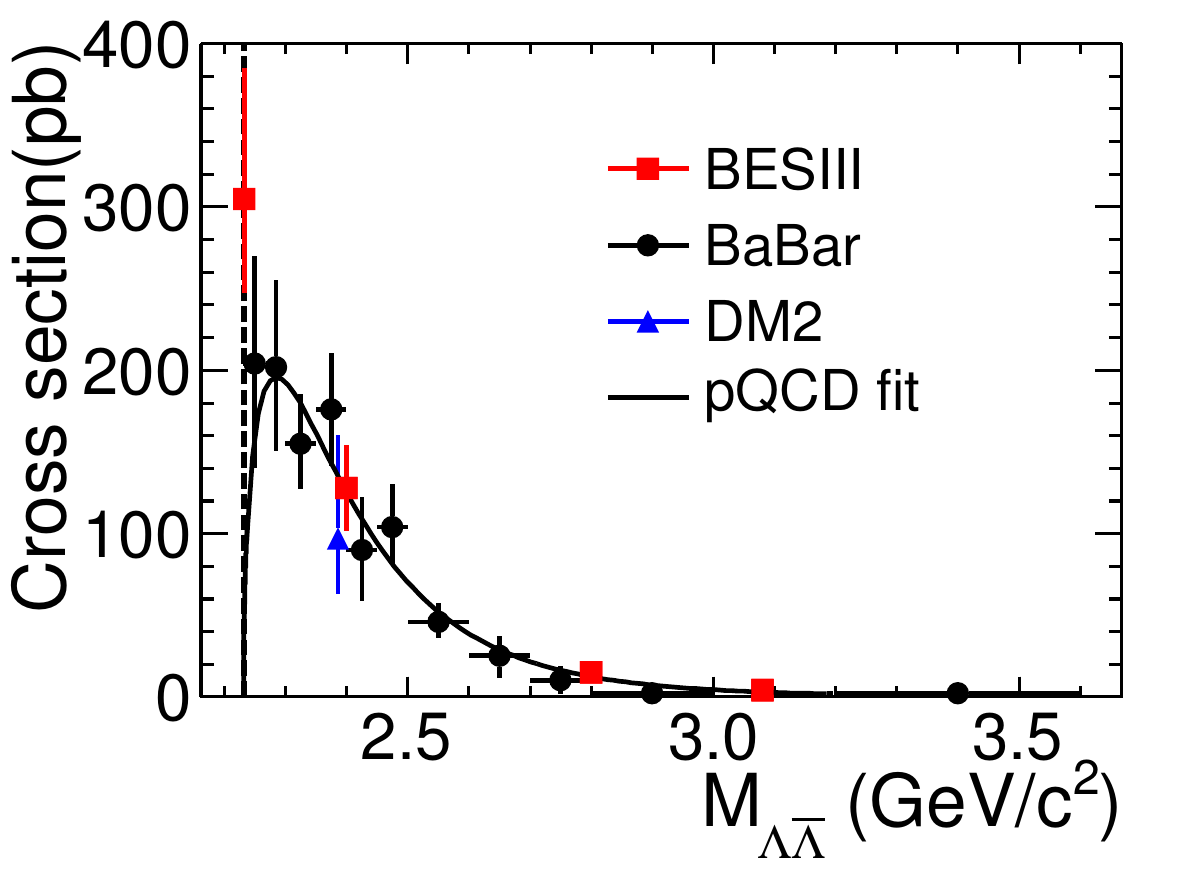}}
\caption[]{Cross section of $e^+e^{-} \rightarrow \Lambda \bar{\Lambda}$.}
\label{lambdaXS}
\end{figure}

The $\Lambda$ was measured previously by the DM2~\cite{DM2BB} and
BaBar~\cite{LamSigBaBar2007} experiments.
BESIII studied the channel $e^+e^{-} \rightarrow
\Lambda \bar{\Lambda}$~\cite{Lambdapair} with a $40.5~\mathrm{pb}^{-1}$ 
data sample collected at four different energy scan points. 
The lowest energy point is 2.2324 GeV, only 1 MeV
above the $\Lambda\bar{\Lambda}$-threshold. 
These data made it possible to measure the Born cross section 
very near threshold. To use the data as efficiently as possible, 
both events where $\Lambda$ and $\bar{\Lambda}$ decayed to 
the charged mode ($\mathrm{Br}(\Lambda \rightarrow p\pi^-) = 64\%$)
and events where the $\bar{\Lambda}$ decayed to the neutral mode
($\mathrm{Br}(\bar{\Lambda} \rightarrow \bar{n} \pi^0) = 36\%$) were
selected. In the first case, the identification relied on finding two
mono-energetic charged pions with evidence for a $\bar{p}$-annihilation
in the material of the beam pipe or the inner wall of the tracking chamber. 
In the second case, the $\bar{n}$-annihilation was identified with a
multi-variate analysis of variables provided by the electromagnetic
calorimeter.  Additonally, a mono-energetic $\pi^0$ was reconstructed 
to fully identify this decay channel. 
For the higher energy points, only the charged decay modes of 
$\Lambda$ and $\bar{\Lambda}$ were reconstructed by
identifying all the charged tracks and using the event kinematics.
The resulting measurement~\cite{Lambdapair} of the Born cross section 
are shown in Fig.~\ref{lambdaXS} together with previous 
measurements~\cite{DM2BB, LamSigBaBar2007}.
The Born cross section near threshold is found to be 
$312 \pm 51(\rm stat.) ^{+72}_{-45}$(\rm sys.) pb.
This result confirms BaBar's measurement~\cite{LamSigBaBar2007} 
but with much higher momentum transfer squared accuracy. 
Since the Coulomb factor is equal to 1 for neutral baryon pairs, 
the cross section is expected to go to zero at threshold. 
Therefore the observed threshold enhancement implies the existence 
of a complicated underlying physics scenario. 
The unexpected features of baryon pair production near threshold have driven 
a lot of theoretical studies, including scenarios that invoke bound states 
or unobserved meson resonances~\cite{Dalkarov, TheoRes, Xiao:2019qhl}. 
It was also interpreted as an attractive Coulomb interaction 
on the constituent quark level~\cite{Resumfct1, Resumfct2}. 
Another possible explanation is the final-state interactions which 
play an important role near the threshold~\cite{FSinter1, FSinter2, FSinter3}.

The spike and the non zero cross section are a big surprise, 
since this cross section was expected to be vanishing, 
because of the phase space factor, 
the baryon center of mass (c.m.) velocity $\beta$. 
It is the same for any neutral particle pair production at threshold.
On the contrary a jump is expected in the case of charged fermion, 
lepton or baryon, pairs. 
In fact the final state long range Coulomb interaction
introduces in the cross section a factor proportional to $\beta^{-1}$, 
which cancels the phase space factor $\beta$. 
This argument is exploited in some more detail in the following.

In the case of a lepton pair an additional final state Coulomb factor 
$\mathcal{C}$ is predicted~\cite{Sommerfeld}, the so called 
Sommerfeld-Schwinger-Sakharov rescattering formula. 
This factor has a very weak dependence on the fermion pair total spin, 
hence it is assumed to be the same for form factors (FF) $G_E$ and $G_M$. 
With a good, non-relativistic, approximation it is:
\begin{eqnarray}
\mathcal{C} = \frac{\pi \alpha}{\beta} \, \frac{1}{1-\exp(-\pi \alpha/\beta)}\, .
\nonumber\end{eqnarray}
There are various relativistic small improvements of this formula, like the substitution 
$\beta \to \tilde\beta= \beta/(1-\beta)$,  which  has been applied in the following. 
This $\mathcal{C}$ factor can also be obtained introducing the Coulomb final state interaction
as the wave function at the origin squared in  $p\overline{p}$ scattering.
Therefore such a Coulomb factor should affect the S wave. 

Essentially the $\beta^{-1}$ divergency comes from the one photon exchange
and the so called resummation factor $\mathcal{R} =
\big[1-\exp(-\pi\alpha/\tilde\beta)\big]^{-1}$ takes into account the many
photons exchange.  The resummation factor for pointlike fermions is so that
the phase space $\beta$ and the corresponding fast increase in the cross
section are restored after few MeV.  Because of the finite colliding beams
energy spread the Coulomb steep rise is hardly seen in the case of lepton
pairs, like $\mu^+\mu^-$ or $\tau^+ \tau^-$.

Hence the clear evidence of a jump in the case of 
{\ensuremath{e^+ e^- \!\rightarrow p\overline{p}}}, followed by a flat
cross section up to about 2100 MeV in the c.m., is very consistent with the
absence of an electromagnetic $\mathcal{R}$ in the case of a charged baryon
pair production.  In fact in this case also strong interactions, that is
gluon exchange, must contribute to $\mathcal{R}$.  Assuming $\mathcal{R}_s =
\big[1-\exp(-\pi \alpha_s/\tilde \beta)\big]^{-1}$, with $\alpha_s$ about
0.5, the flat {\ensuremath{e^+ e^- \!\rightarrow p\overline{p}}} 
cross section on a hundred MeV scale is well reproduced. 
By the way S wave dominance at threshold is required also by analyticity 
and, as a consequence, there should be one FF, since both imply that 
$G_E^p(4 M_p^2)=G_M^p(4 M_p^2)$, at least close to the threshold.

For more than 30 years to get the proton FF the Coulomb factor for pointlike
fermions has been applied.  Therefore the proton FF, being obtained from a
flat cross section divided by a steep increasing factor, has shown an
apparent steep decrease, simulating the tail of a narrow resonance below
threshold.  Of course, assuming the FF are
defined as additional factors with respect to the pointlike amplitude, one
could take into account the pointlike Coulomb factor, but it seems unlike to
attach a physical meaning to the sharp decrease at threshold obtained in
this way.  To avoid this kind of ambiguity a better definition of FF in the
time-like region might be proposed.
 
It has also been argued~\cite{Rinaldo} that for the proton FF at threshold verify
the identity $G_E^p(4 M_p^2)=G^p_M(4 M_p^2)= 1$, as achieved if the flat
cross section is extrapolated down to the threshold and the divergent factor
$\pi \alpha/\beta$ is applied.  After all such a result should not be so
unexpected, since at threshold the overlap between $\overline{p}$ and $p$ wave
function looks like the overlap of initial and final proton wave function in
the case of electron proton scattering at a vanishing electron energy. 

Coming back to the case of neutral baryon pair production, the fact that the
cross section {\ensuremath{e^+ e^- \!\rightarrow \Lambda\overline{\Lambda}}} 
is non zero at threshold strongly suggests that it is
due to an unexpected Coulomb interaction at the quark level!  This is
surprising, since Coulomb interaction is a long range one, while strong
interactions have a short range.  As a consequence it was assumed until now
that on a short time scale the hadron pair is created and after, on a much
longer time scale, the Coulomb interaction takes over.  To evaluate the
expected cross section at threshold at the quark level a naive prediction
(formulated some years ago~\cite{Rinaldo}) would be that this cross section 
scales as the sum of the valence quarks charge squared, namely:
\begin{eqnarray}
\sigma(4 M_\Lambda^2)= \sigma(4 M_p^2) \times \left(\frac{M_p}{M_\Lambda}\right)^2\times \frac{Q_u^2+Q_d^2+Q_s^2}{Q_u^2+Q_u^2+Q_d^2}= 400 \,{\rm pb}\,,
\nonumber\end{eqnarray}
where $Q_q$ is the charge of the quark $q$.
Such expectation would be consistent with $G_E^p(4 M_p^2)=G_M^p(4 M_p^2)= 1$ 
in the proton case, since $Q_u^2+Q_u^2+Q_d^2 = 1$.
Surprisingly enough this very naive prediction is in good agreement with the 
{\ensuremath{e^+ e^- \!\rightarrow \Lambda\overline{\Lambda}}} experimental result. 

However this Coulomb contribution for neutral baryon pair production should
vanish soon as the c.m. energy increased, somewhat in agreement with the
aforementioned expectation as well as with the already mentioned 
$B${\scriptsize $A$}$B${\scriptsize $AR$} cross section 
$\sigma= 200\pm 60 \pm 20$~pb, which is an average value from
threshold up to $W= 2270$ MeV~\cite{LamSigBaBar2007}.
Such a behaviour can be reproduced, if attractive and repulsive Coulomb
factors both are taken into account with their proper sign in the amplitude,
namely:
\begin{eqnarray}
\mathcal{R}_{\rm eff} = \left[\frac{1}{\sqrt{\exp(\pi\alpha_{\rm eff}/\tilde\beta)-1}}- \frac{1}{\sqrt{1-\exp(-\pi \alpha_{\rm eff}/\tilde\beta)}}\right]^2
\,.
\nonumber\end{eqnarray}
In this way, depending on $\alpha_{\rm eff}$, as anticipated, since the
hadron is neutral, they cancel each other except close to the threshold,
producing a cusp in the cross section.  Of course a further strong
interaction contribution has to be included, which should behaves like:
$\beta \times G(W^2)$, where $G(W^2)$ should take into account the tails of
resonances below threshold and asymptotically should scale like
$G(W^2)\propto W^{-10}$.  If Coulomb interaction is limited to a cusp it
might be meaningful to disentangle the two contributions.

To trust all these speculations, confirmations are very welcome.
More measurements will be needed in a range just above threshold 
to check this cusp behaviour 
and to check if indeed $\alpha \ll \alpha_{\rm eff} \ll \alpha_s$, 
assuming the model is meaningful.
Furthermore, to establish a theory, a measurement of other 
hyperon cross sections at their own threshold will be needed too. 
Assuming the other charged and neutral hyperons have a behaviour 
similar to {\ensuremath{e^+ e^- \!\rightarrow p\overline{p}}} or 
{\ensuremath{e^+ e^- \!\rightarrow \Lambda\overline{\Lambda}}}, 
a series of spikes and jumps is expected. 

\begin{figure}[htbp]
\includegraphics[width=0.49\textwidth]{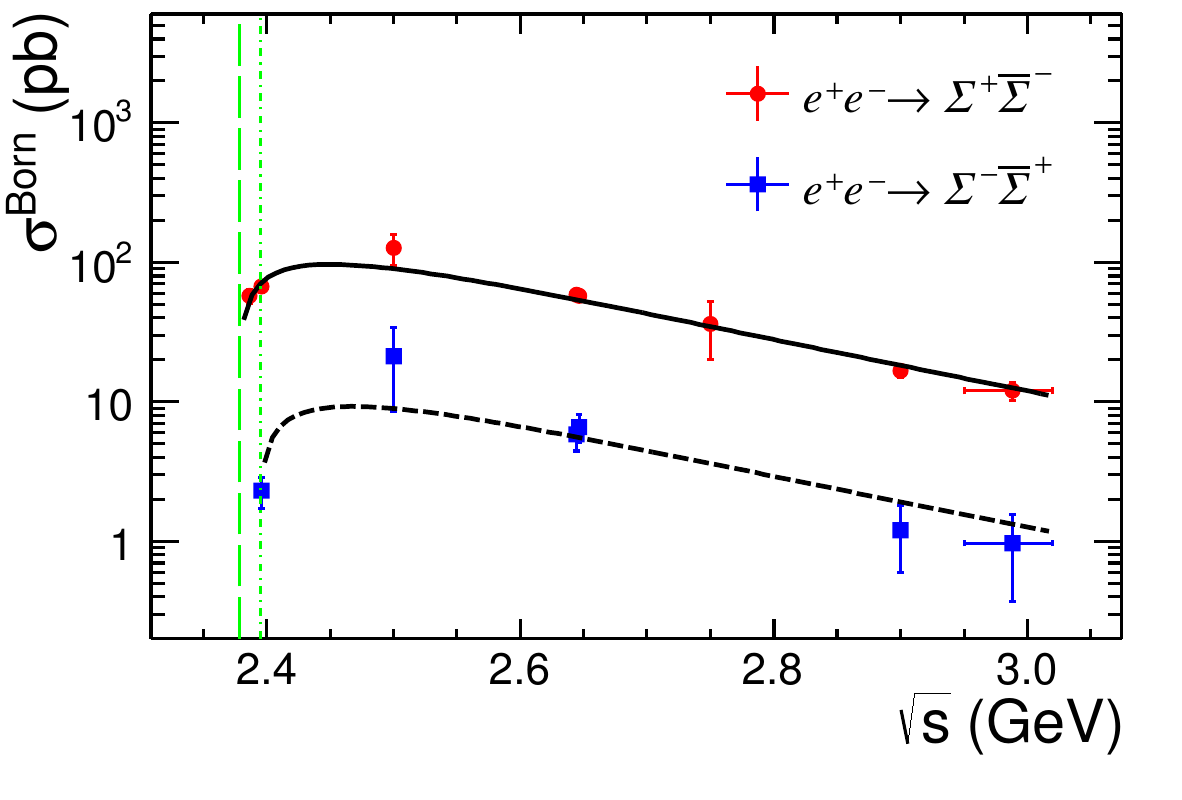}
\hspace{0.5cm}
\includegraphics[width=0.49\textwidth]{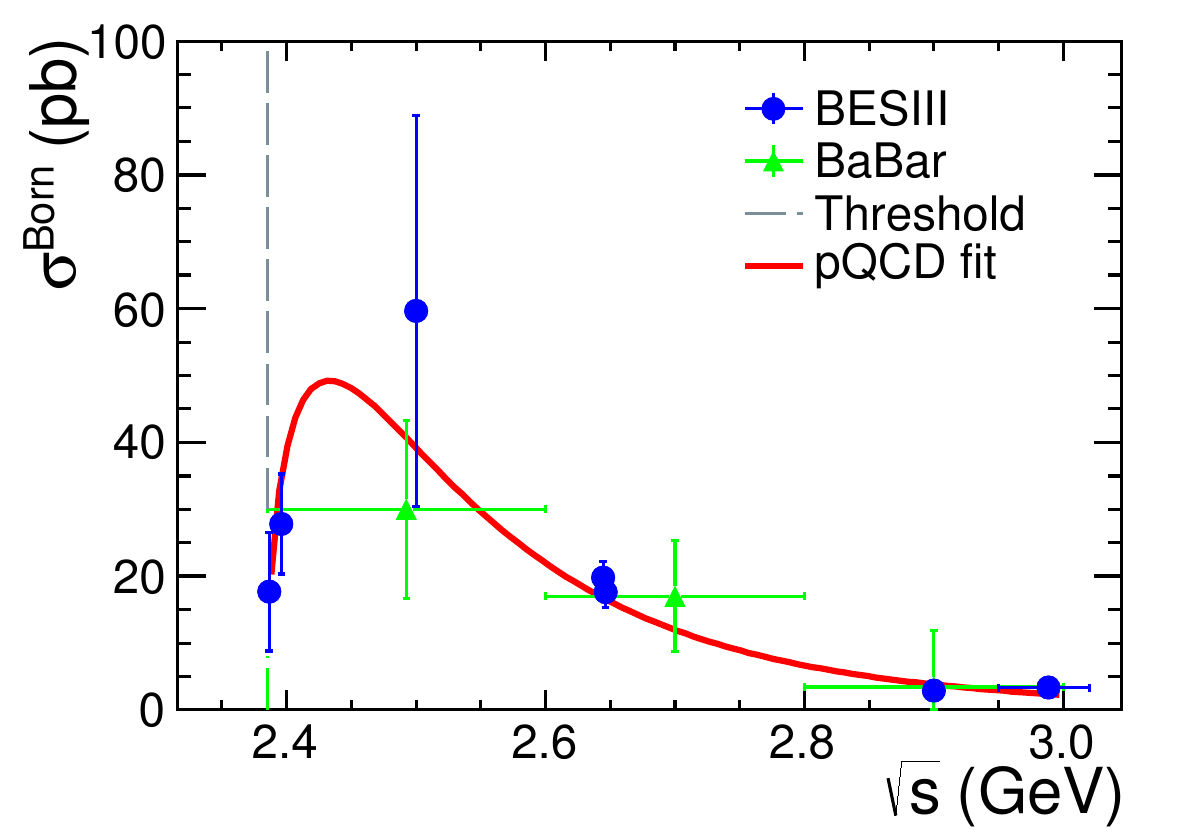}
\caption{Left: the cross section lineshapes for 
$e^{+}e^{-}\to\Sigma^{+}\bar{\Sigma}^{-}$ reactions (circles) and 
$e^{+}e^{-}\to\Sigma^{-}\bar{\Sigma}^{+}$ (squares)~\cite{SigmaPLB2021}. 
The solid and dashed smooth lines are the pQCD fits.
The vertical lines denoted their production thresholds.
Right: Comparison plot of the cross sections for 
$e^{+}e^{-}\to\Sigma^{0}\bar{\Sigma}^{0}$ reaction. 
The triangles in green are results from BaBar~\cite{LamSigBaBar2007}. 
The solid line in red shows the pQCD fit.}
\label{SigmaXSfit}
\end{figure}

Such that, the cross sections near thresholds have been measured 
by BESIII for other baryon pairs, for examples singly-stranged 
$\Sigma^{+}\bar{\Sigma}^{-}$, 
$\Sigma^{-}\bar{\Sigma}^{+}$~\cite{SigmaPLB2021} and 
$\Sigma^{0}\bar{\Sigma}^{0}$~\cite{Sigma0PLB2022}
as shown in Fig.~\ref{SigmaXSfit}, 
doubly-stranged $\Xi^{-}\bar\Xi^{+}$~\cite{chargedXi} and 
$\Xi^{0}\bar\Xi^{0}$~\cite{neutralXi} 
as shown in Fig.~\ref{XiXiGeffXS}.

\begin{figure}[htbp]
\begin{center}
\includegraphics[width=0.9\textwidth]{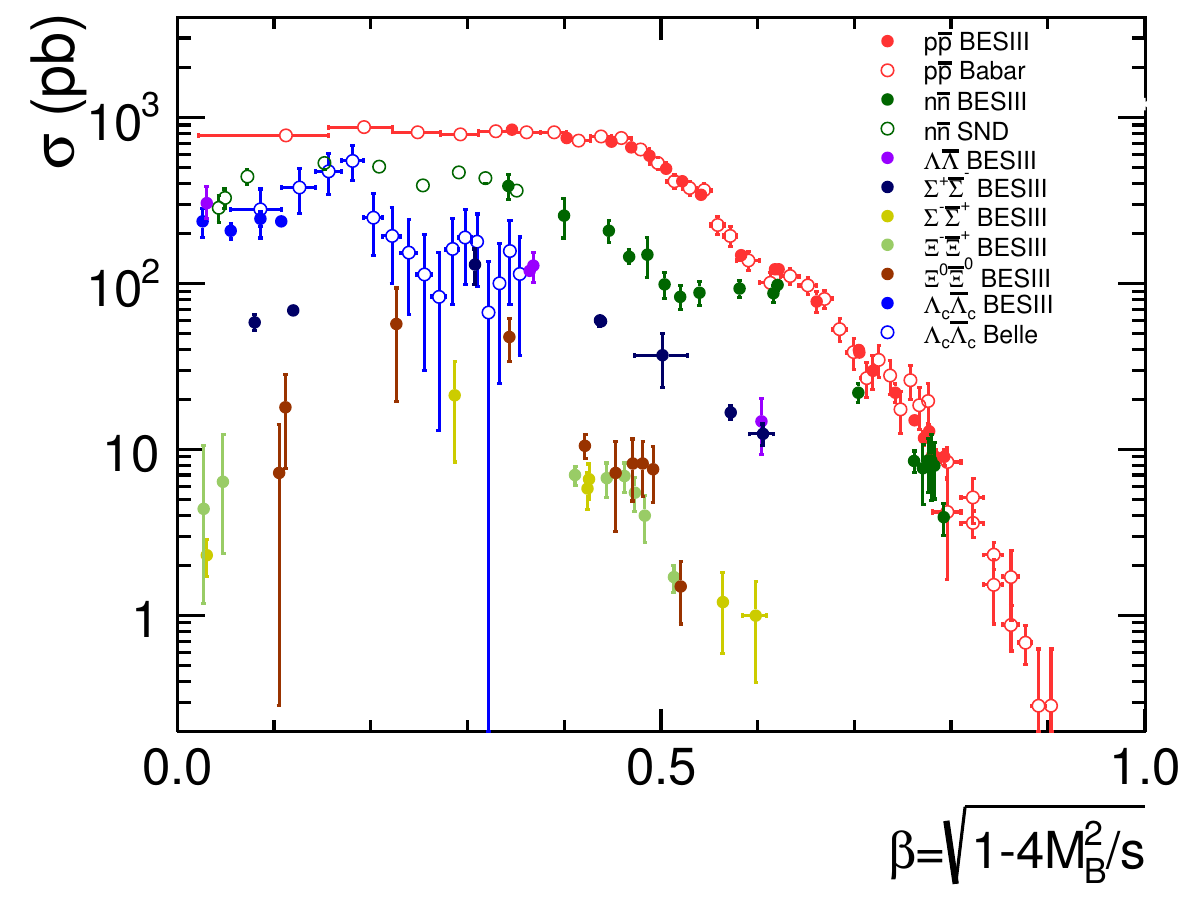}
\caption{Cross sections of $B \bar{B}$ pairs measured so far: 
$p\bar{p}$ by BaBar~\cite{ppbarBaBar1,ppbarBaBar2} and BESIII~\cite{ppbar2020}, 
$n\bar{n}$ by SND~\cite{SND2014, SNDnn} and BESIII~\cite{nnbar},
$\Lambda\bar{\Lambda}$ by BESIII~\cite{Lambdapair},
$\Sigma^{+}\bar{\Sigma}^{-}$/$\Sigma^{-}\bar{\Sigma}^{+}$ 
by BESIII~\cite{SigmaPLB2021},
$\Xi^{-}\bar{\Xi}^{+}$/$\Xi^{0}\bar{\Xi}^{0}$ 
by BESIII~\cite{chargedXi, neutralXi},
$\Lambda_c^+\bar{\Lambda}_c^-$ by Belle~\cite{LambdacBelle} 
and BESIII~\cite{Lambdacpair}. Plot is from~\cite{NSR2021}.}
\label{xs_vs_beta}
\end{center}
\end{figure}

Figure~\ref{xs_vs_beta} shows the cross section lineshapes for 
a variety of baryon-antibaryon pairs have been measured so far~\cite{NSR2021}.
They all seem to share the common feature with a plateau 
starting from the baryon-pair production threshold,
though for some channels ideally more statistics are needed.

\section{Hyperon form factors}

In the time-like region ($q^2 > 4m_B^2 > 0$, $B$ refers to baryon), the form factors $G_E(q²)$ and $G_M(q²)$  are complex functions, $G_E(q²)=|G_E(q²)|e^{i\Phi_E}$ and $G_M(q²)=|G_M(q²)|e^{i\Phi_M}$ with a relative phase $\phi=\Delta\Phi=\Phi_M-\Phi_E$.
The effective form factor, as defined for spin $\frac{1}{2}$ baryons
\begin{equation}
|F|^2=\frac{2\tau|G_M|^2+|G_E|^2}{2\tau+1}=\frac{2\tau}{(2\tau+1)}\frac{3q^2\sigma}{4\pi\alpha^2\beta}
\label{eq:eff}
\end{equation}
is obtained in the same way as for nucleons, \textit{i.e.} from the energy dependence of the total cross section. 
Note that $\tau=q^2/(4m^2)$ and $\beta=1-1/\tau$. 
The electric and the magnetic form factors can then be expressed in the following way:

\begin{equation}
|G_M|^2=\frac{2\tau +1}{2\tau+R^2}|F|^2
\label{eq:sachsm}
\end{equation}
\begin{equation}
|G_E|^2=R^2\frac{2\tau +1}{2\tau+R^2}|F|^2.
\label{eq:sachse}
\end{equation}

The ratio $R=|G_E(q²)|/|G_M(q²)|$ can be extracted from the scatering angle of the outgoing baryon:
\begin{equation}
\frac{d\sigma}{d\cos\theta} = N_1((1+\cos^2\theta)|G_M|^2+\frac{1}{\tau}(1-\cos^2\theta)|G_E|^2)
\label{eq:diff}
\end{equation}
where $N_1$ is an proportionality factor, constant in the angle and which depends on $\alpha$, $\beta$, the Coloumb correction factor $C$ and $q$. 

\begin{figure}[htbp]
  \centering
  \includegraphics[width=0.95\textwidth]{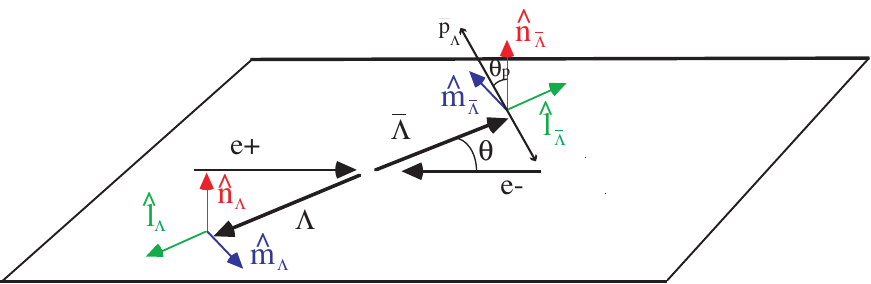}
\caption{The coordinate system of the $e^{+}e^{-} \rightarrow Y\overline{Y}$ process}
  \label{fig:coord}
\end{figure}

Now consider production of spin $\frac{1}{2}$ hyperon $Y$, produced in $e^+e^- \rightarrow Y\overline{Y}$ and decaying into a spin $\frac{1}{2}$ baryon and a pseudoscalar meson, for example $e^+e^- \rightarrow \Lambda\overline{\Lambda}, \Lambda \rightarrow p \pi^-$. The reference system is defined in Fig. \ref{fig:coord}. The scattering plane is spanned by the $e^+$ and the outgoing hyperon, \textit{i.e.} its normal is defined by $\hat{n}=\hat{e_{e^+}} \times \hat{e_{\Lambda}}$ or equivalently, by the $e^-$ and the outgoing antihyperon. 
The angle $\theta$ is the scattering angle of the hyperon/antihyperon and the angle between the decay proton(antiproton) and $\hat{n}$ is $\theta_p$($\theta_{\overline{p}}$). 

The relative phase between the form factors has a polarizing effect on the final state hyperons. In Ref. \cite{dubnicka}, the polarization in terms of electromagnetic form factors $G_E(q^2)$ and $G_M(q^2)$ has been derived for unpolarised $e^{+}$ and $e^{-}$, resulting in 

\begin{equation}
P_n=- \frac{\sin2\theta Im[G_E(q^2)G_M^*(q^2)]/\sqrt{\tau}}{(|G_E(q^2)|^2\sin^2\theta)/\tau+|G_M(q^2)|^2(1+\cos^2\theta)}= - \frac{\sin(2\theta) \, \sin(\phi) /\sqrt{\tau}}{\frac{R \, \sin^2(\theta)}{\tau} + \frac{1+\cos^2(\theta)}{R}}
\label{eq:polee}
\end{equation}
\vspace{5mm}

where $Im[G_E(q^2)G_M^*(q^2)]$ = $|G_E(q^2)||G_M(q^2)|\sin\phi$ and $R=|G_E(q²)|/|G_M(q²)|$. $\phi$ is the relative phase between the electric and the magnetic form factor and $\tau=q^2/(4m_Y^2)$. Thus, a measurement of the polarization determines the modulus of $\sin\phi$. From Eq. \ref{eq:polee} it is clear that the polarization strongly depends on the scattering angle $\theta$. This has to be taken into account when extracting the phase.

In a similar way, the real part of the $G_EG_M^*$, $Re[G_E(q^2)G_M^*(q^2)]$ = $|G_E(q^2)||G_M(q^2)|\cos\phi$ can be obtained from the correlation of the $\Lambda$ and $\overline{\Lambda}$ spins in the $\hat{m}$ and $\hat{l}$ directions  \cite{dubnicka}:

\begin{equation}
C_{lm}=\frac{\sin2\theta Re[G_E(q^2)G_M^*(q^2)]/\sqrt{\tau}}{(|G_E(q^2)|^2\sin^2\theta)/\tau+|G_M(q^2)|^2(1+\cos^2\theta)}
\label{eq:corree}
\end{equation}
\vspace{5mm}

Thus, by measuring the effective form factor, the angular distribution of the hyperon and the polarisation and $\Lambda\overline{\Lambda}$ spin correlations, the time-like form factors can be fully determined.

\subsection{Existing data and theoretical predictions}

Very few measurements have been performed on $e^{+}e^{-} \rightarrow
Y\overline{Y}$ channels.  The DM2 collaboration measured the $e^{+}e^{-}
\rightarrow \Lambda\overline{\Lambda}$ cross section at a CM energy 2.386
GeV \cite{dm2} and found it to be $\sigma(e^{+}e^{-} \rightarrow
\Lambda\overline{\Lambda})=100_{-35}^{+65}$ $pb$.

BABAR used ISR data at a $e^{+}e^{-}$ CM energy of 10.58 GeV to extract the
cross section/effective form factor 12 points in
$M(\Lambda\overline{\Lambda})$ \cite{LamSigBaBar2007}. Their total event
sample consisted of only $\approx$200 events and the relative error in each
point was large (typically $>30\%$). The ratio was extracted in two different
energy ranges:
\begin{itemize}
\item $|G_E/G_M| = 1.73_ {-0.57}^{+0.99}$ for 2.23 $< q <$ 2.4 GeV
\item $|G_E/G_M| = 0.71_ {-0.71}^{+0.66}$ for 2.24 $< q <$ 2.6 GeV
\end{itemize}

They also calculated the phase, but integrated over all scattering angles
and all energies, and assuming a ratio of $|G_E/G_M| = 1$.  Consequently 
the result was inconclusive: $-0.76 < \sin\phi < 0.98$.

On the theoretical side, Czyz, Grzelinska and K\"{u}hn adopted in 
Ref. \cite{czyz} a model from K\"{o}rner and Kuroda \cite{korner} 
which predicts real form factors, \textit{i.e.} zero phase and 
unpolarised hyperons.
Calculations have been made by Bartos \textit{et al.} \cite{bartos} 
of the moduli of the Sachs form factors of $\Lambda$, $\Sigma$ and 
$\Xi$ based on the Unitary and Analytic model \cite{dubnickova}. 
There are recent predictions by the same group of vector- and 
tensor polarisations of nucleons \cite{dubnickahep}.
\begin{itemize}
\item more precise data on hyperon form factors, \textit{i.e.} the effective form factor and the ratio $R$, or 
\item lepton decay widths for the ground state, the first excitation and the second excitaion of $\rho$, $\omega$ and $\phi$ mesons, from which one could determine the universal vector meson coupling constants.
\end{itemize}
More experimental data will therefore stimulate the theoretical activity of the field. 

\subsection{Hyperon form factors with BESIII}

\subsubsection{The $\Lambda$ hyperon}

\begin{figure}[htbp!]
\centerline{\includegraphics[width=0.6\linewidth]{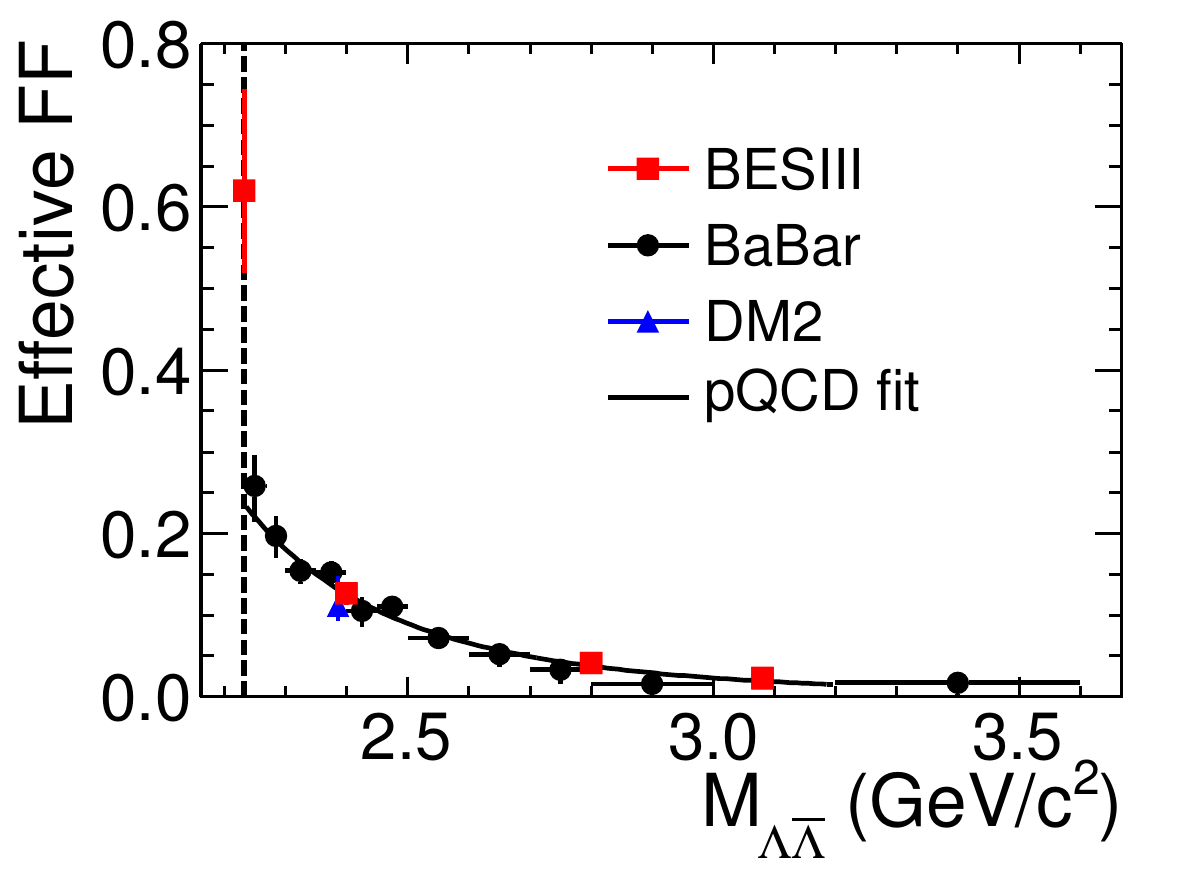}}
\caption[]{$\Lambda$ effective form factor.}
\label{lambdaFF}
\end{figure}

Based on test run data at four energies, the $\Lambda$ 
effective form factor was extracted by BESIII~\cite{Lambdapair},
as shown in Fig.~\ref{lambdaFF}.

\begin{figure}[htbp!]
\begin{center}
\includegraphics[width=0.49\textwidth]{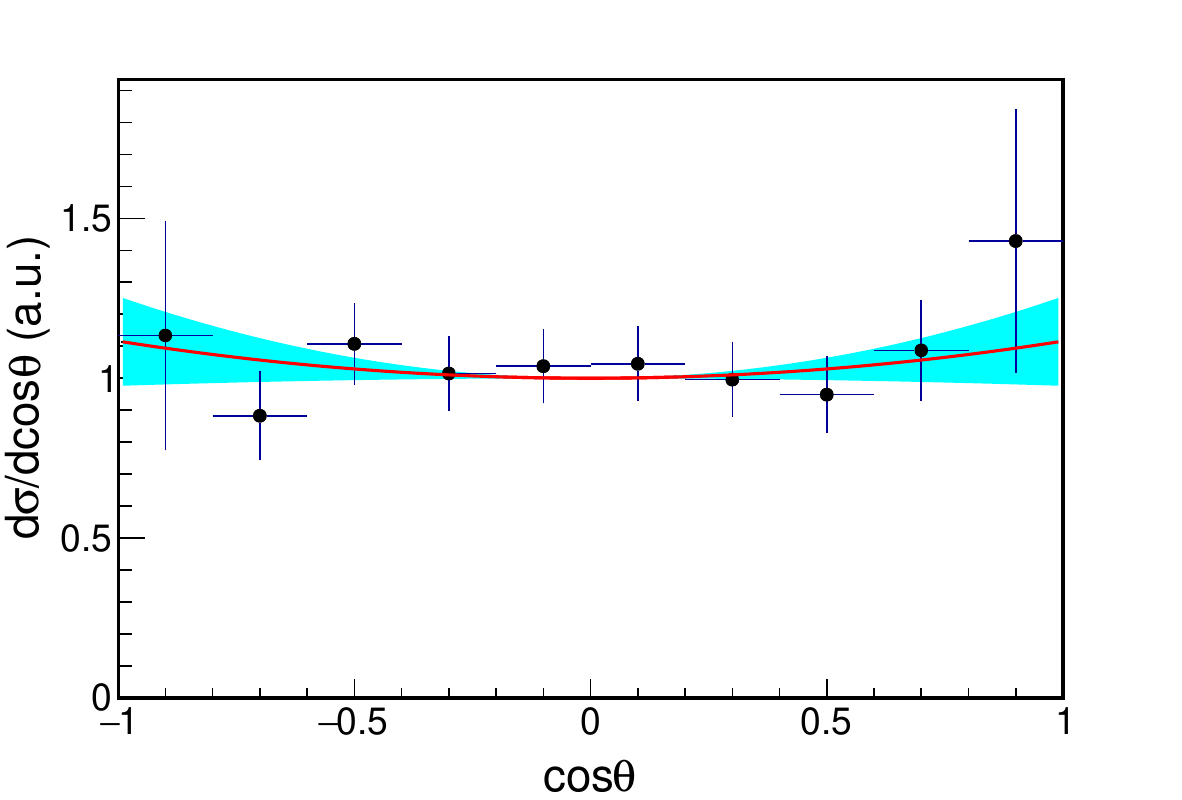}
\put(-190, 120){(a)}
\includegraphics[width=0.49\textwidth]{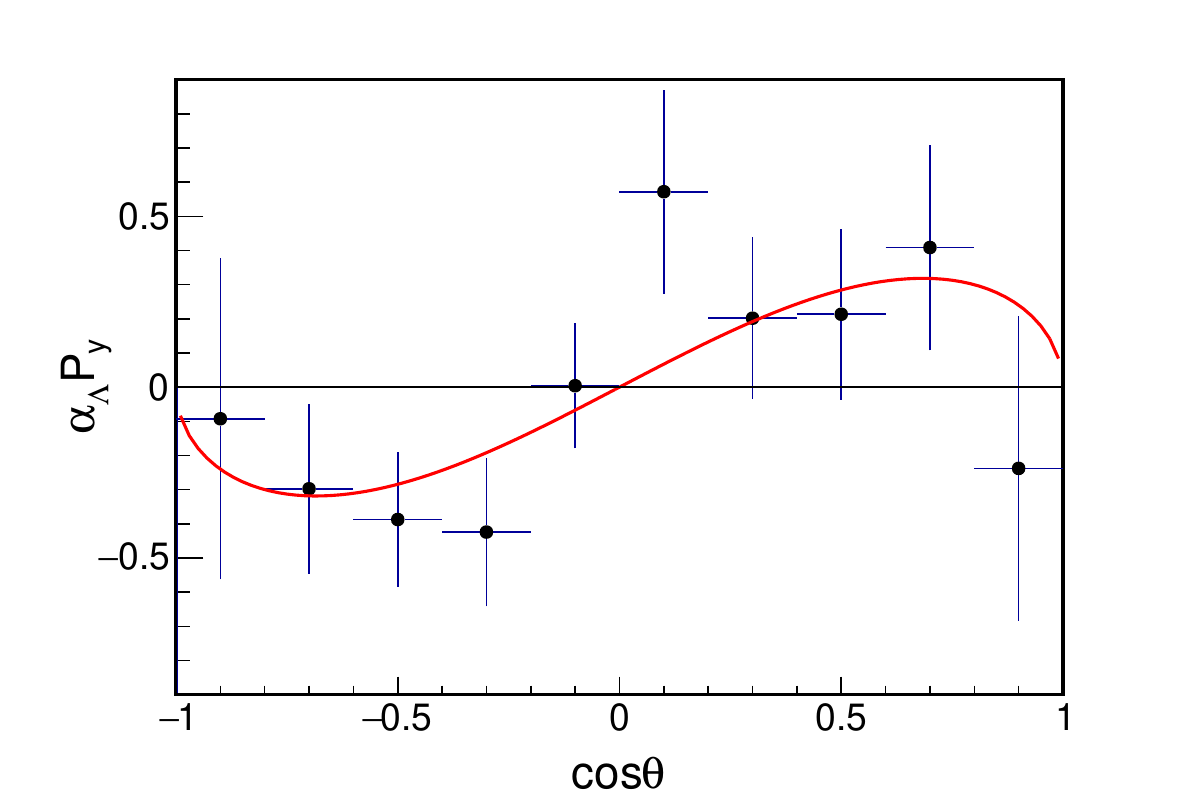}
\put(-180, 120){(b)}
\caption{(a) The acceptance corrected $\Lambda$ scattering angle 
distribution for $e^+e^{-} \rightarrow \Lambda \bar{\Lambda}$ at
$M_{\Lambda \bar{\Lambda}} = 2.396$ GeV.
(b) The product of the $\Lambda$ decay parameter $\alpha_\Lambda$ 
and $\Lambda$ polarization $P_y$ as a function of the scattering angle.}
\label{cosThetaLambda}
\end{center}
\end{figure}

According to the optical theorem, there is a nonzero relative phase
between $G_E$ and $G_M$.
At $M_{\Lambda \bar{\Lambda}} = 2.396$ GeV, where 
the largest $\Lambda \bar{\Lambda}$ sample of 555 events from 
66.9 pb$^{-1}$ data was accumulated later, a multidimensional analysis 
was used to make a full determination of the $\Lambda$ 
electromagnetic form factors for the first time for any baryon; 
the relative phase difference is 
$\Delta\Phi = 37^\circ \pm 12^\circ \pm 6^\circ$~\cite{LambdaPRL2019} 
with the input parameter $\alpha_\Lambda = 0.750\pm0.010$ 
measured from $J/\psi$ decays~\cite{LambdaPolar}.
The improved determination of $\alpha_\Lambda$ also has profound 
implications for the baryon spectrum, since fits to such observables 
by theoretical models are a crucial element in determining 
the light baryon resonance spectrum, which provides a point of 
comparison for theoretical approaches~\cite{alphaLambda}.
The $|G_E/G_M|$ ratio was determined to be 
$R = 0.96 \pm 0.14(\rm stat.) \pm 0.02(\rm sys.)$ and 
the effective form factor at $M_{\Lambda \bar{\Lambda}} = 2.396$ GeV
was determined to be
$|G_{eff}| = 0.123 \pm 0.003 (\rm stat.) \pm 0.003 (\rm sys.)$.
The $\Lambda$ angular distribution and the polarization 
as a function of the scattering angle are shown in 
Fig.~\ref{cosThetaLambda}(a) and (b)~\cite{LambdaPRL2019}, respectively.
This first complete measurement of the hyperon electromagnetic 
form factor is a milestone in the study of hyperon structure.

\subsubsection{The $\Sigma$ hyperons}

BESIII studied the processes $e^+e^-\to\Sigma^{\pm}\bar{\Sigma}^{\mp}$ 
and $e^+e^-\to\Sigma^0\bar{\Sigma}^0$ reactions from 2.3864 to 3.0200 GeV 
and determined the timelike EMFFs of $\Sigma$ hyperons 
with high precision~\cite{SigmaPLB2021, Sigma0PLB2022}. 
Born cross sections of $\Sigma^{\pm}\bar{\Sigma}^{\mp}$ pair productions, 
effective form factors $|G_{\mathrm{eff}}|$ of $\Sigma^{+}$ and $\Sigma^{-}$, 
the ratios of $\Sigma^{+}$ electric and magnetic FFs $|G_{E}/G_{M}|$, 
were reported~\cite{SigmaPLB2021}.  
For c.m.energies near threshold, a novel method was used to reconstruct 
the neutral channel $e^{+}e^{-}\to\Sigma^{0}\bar{\Sigma}^{0}$ 
whereas a single-hyperon-tag method was applied for c.m.~energies 
between 2.5000 and 3.0200 GeV.  
Born cross sections are measured with significantly improved
precision~\cite{Sigma0PLB2022} to those of BaBar~\cite{LamSigBaBar2007}.  
The $|G_{\rm{eff}}|$ of $\Sigma^{0}$ was also reported.  

Near production threshold of $\Sigma^{\pm}\bar{\Sigma}^{\mp}$ pairs, 
the cross section were observed to be $(58.2\pm5.9^{+2.8}_{-2.6})$~pb 
and $(2.3\pm0.5\pm0.3)$~pb, which disagrees with the pointlike expectations
close to threshold of $848(m_{p}/m_{B})^{2}$~pb. 
The cross section lineshapes presented in Fig.~\ref{SigmaXSfit} for
$e^{+}e^{-}\to\Sigma^{+}\bar{\Sigma}^{-}$ and
$e^{+}e^{-}\to\Sigma^{-}\bar{\Sigma}^{+}$ 
are well-described by pQCD-motivated functions. 
The ratio of the
$\sigma^{\text{Born}}(e^{+}e^{-}\to\Sigma^{+}\bar{\Sigma}^{-})$ to
$\sigma^{\text{Born}}(e^{+}e^{-}\to\Sigma^{-}\bar{\Sigma}^{+})$ 
was found to be $9.7\pm1.3$, which is inconsistent with predictions.
The EFF is proportional to the square root of the cross section, 
and the observed ratio of
$|G_{\mathrm{eff}}^{\Sigma^{+}}(s)|/|G_{\mathrm{eff}}^{\Sigma^{-}}(s)|$ 
was found to be consistent with 3, which is the ratio of 
the incoherent sum of the squared charges of valence quarks in 
$\Sigma^{+}$ and $\Sigma^{-}$ baryons, $\sum_{q} Q_{q}^{2}$, with $q=u,d,s$. 
Furthermore, the EMFF ratio $|G_{E}(s)/G_{M}(s)|$ of the $\Sigma^{+}$ 
was reported through an angular analysis at three high-statistics 
energy points, 2.3960, 2.6444, and 2.6464 GeV for 
$e^{+}e^{-}\to\Sigma^{+}\bar{\Sigma}^{-}$. 
Based on the polar angular distribution of $\Sigma^{+}$, 
the ratio $|G_{E}(s)/G_{M}(s)|$ of the $\Sigma^{+}$ baryon 
was determined to be $|G_{E}(s)/G_{M}(s)| =1.83\pm 0.26$ near threshold,
which is significantly higher than 1~\cite{SigmaPLB2021}. 

\begin{figure}[htbp]
\begin{center}
\includegraphics[width=0.6\textwidth]{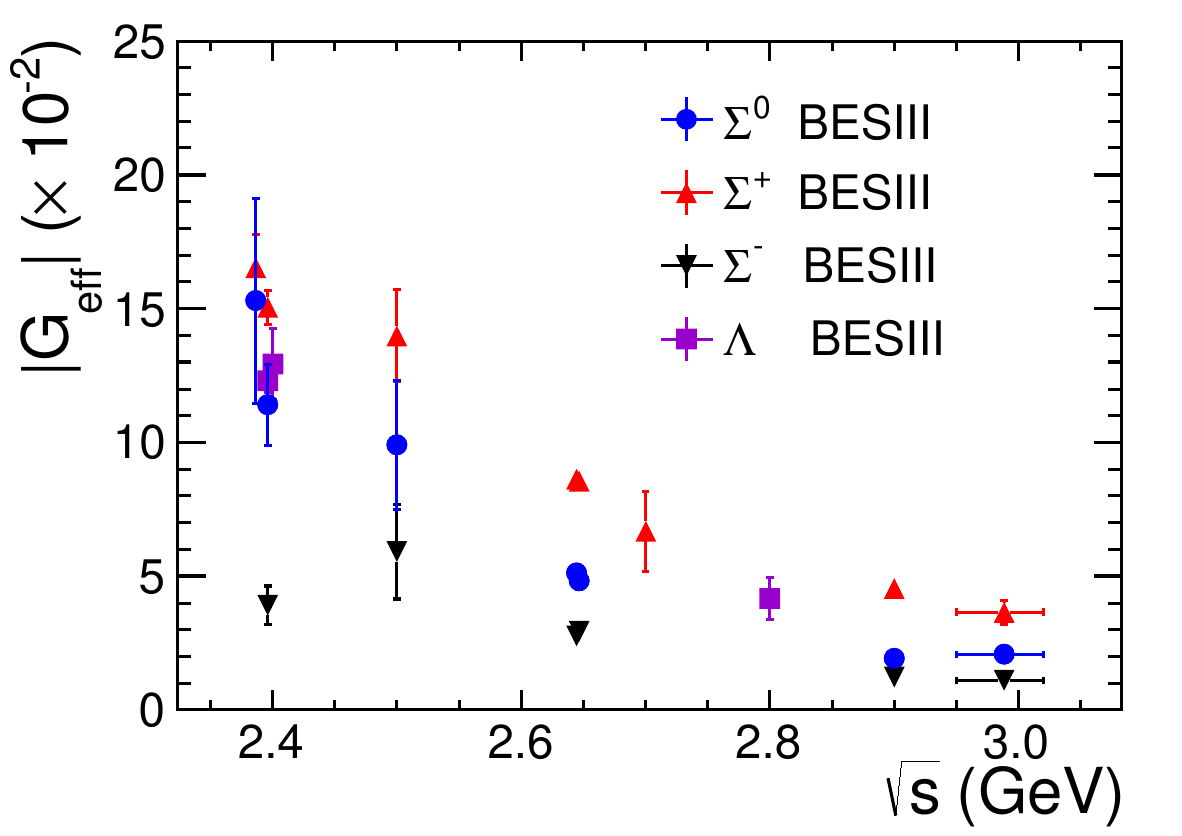}
\caption{Effective form factors of hyperons measured at 
BESIII~\cite{SigmaPLB2021, Sigma0PLB2022, Lambdapair, LambdaPRL2019}.}
\label{SigmaGeff}
\end{center}
\end{figure}

\subsubsection{The $\Xi$ hyperons}
\begin{figure}[htbp]
\begin{center}
\includegraphics[width=0.5\textwidth]{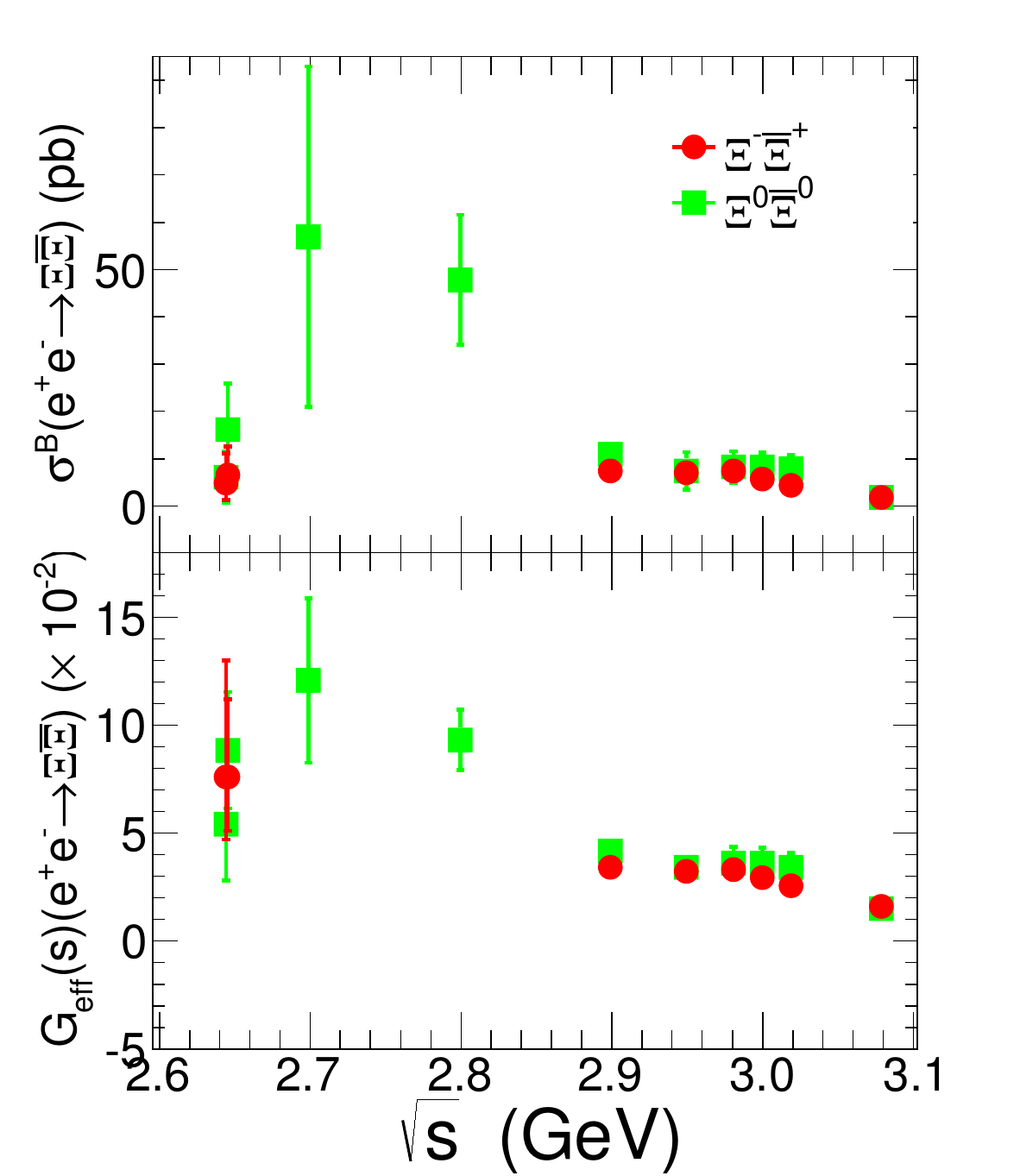}
\caption{Born cross sections (\textbf{top}) and EFFs (\textbf{bottom}) 
of $e^{+}e^{-}\rightarrow\Xi^{-}\bar\Xi^{+}$ and 
$e^{+}e^{-}\rightarrow\Xi^{0}\bar\Xi^{0}$ from 2.6 to 3.1 GeV.}
\label{XiXiGeffXS}
\end{center}
\end{figure}

BESIII measured Born cross sections and EFFs for the processes 
$e^{+}e^{-}\rightarrow\Xi^{-}\bar\Xi^{+}$~\cite{chargedXi} and 
$e^{+}e^{-}\rightarrow\Xi^{0}\bar\Xi^{0}$~\cite{neutralXi} 
based on a single hyperon tag method using data collected 
at c.m. energies between 2.644 and 3.080 GeV.
Figure~\ref{XiXiGeffXS} shows the measured Born cross sections 
and EFFs for the two processes.

\section{Form factors of charmed baryon}
\begin{figure}[t]
\begin{center}
\begin{minipage}{0.45\linewidth}
\centerline{\includegraphics[width=1\linewidth,height=0.75\linewidth]{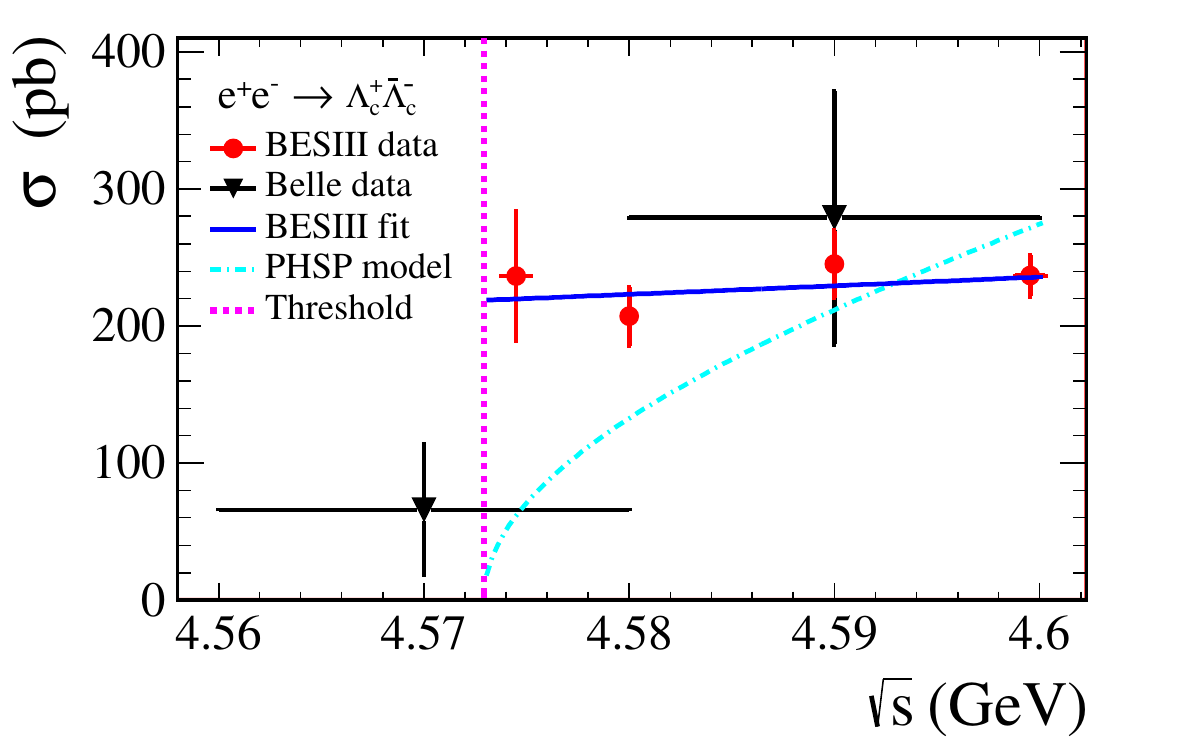}}
\end{minipage}
\put(-45,55){(a)}
\begin{minipage}{0.46\linewidth}
\centerline{\includegraphics[width=1\linewidth,height=0.75\linewidth]{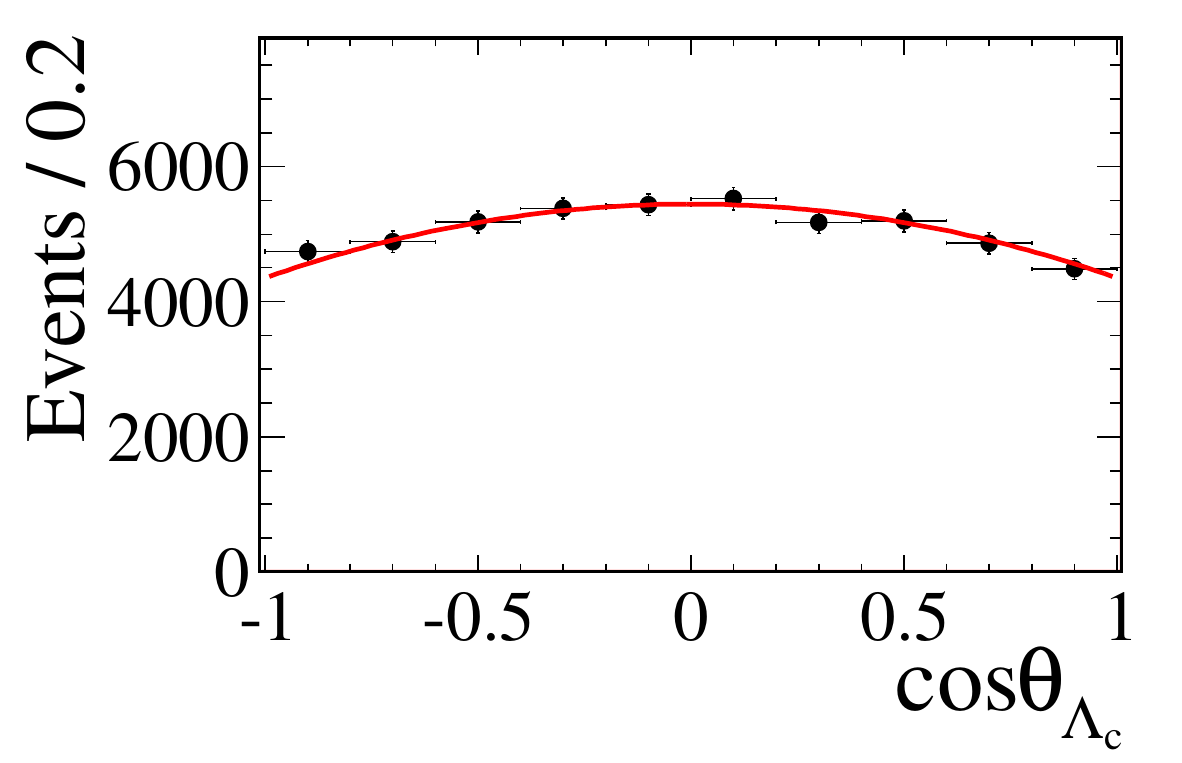}}
\end{minipage}
\put(-40,55){(b)}
\end{center}
\caption[]{The Born cross section of $e^+e^- \to \Lambda_c^+ \bar{\Lambda}_c^-$
obtained by BESIII and Belle (a).
The angular distribution and corresponding fit results in data at
$\sqrt{s} = 4.5995$ GeV (b).}
\label{lambdac2}
\end{figure}

Higher energy opens a windows for charmed baryon studies.
Previously there were no sufficient data above charmed baryon
production threshold, and therefore no form factor results were reported.
The only measurement of cross section of the process 
$e^+e^- \to \Lambda_c^+ \bar{\Lambda}_c^-$
is from the Belle experiment, which measured the cross section 
using ISR technique~\cite{LambdacBelle}, and reported a lineshape 
that implied the existence of a likely resonance, called the Y(4660).
Based on $631.3~\mathrm{pb}^{-1}$ data collected in 2014 at four 
energy points $\sqrt{s} = $4.5745, 4.5809, 4.5900 and 4.5995 GeV,
BESIII measured the $\Lambda_c^+ \bar{\Lambda}_c^-$ cross section
with unprecedented precision~\cite{Lambdacpair}.
The lowest energy point is only 1.6 MeV above the
$\Lambda_c^+ \bar{\Lambda}_c^-$ threshold.
At each of the energy points, ten Cabibbo-favored hadronic decay modes,
$\Lambda_c^+ \rightarrow p K^-\pi^+$, $p K_S^0$, $\Lambda \pi^+$,
$p K^-\pi^+ \pi^0$, $p K^0 \pi^0$, $\Lambda \pi^+ \pi^0$,
$p K_S \pi^+ \pi^-$, $\Lambda \pi^+ \pi^+ \pi^-$, $\Sigma^0 \pi^+$,
and $\Sigma^+\pi^+\pi^-$, as well as the corresponding charge-conjugate
modes were studied.
The total Born cross section is obtained from the weighted average
of the 20 individual measurements, and
the results are shown in Fig.~\ref{lambdac2}(a). 
Similar to the case for $e^+e^- \to p \bar{p}$, an abrupt rise 
in the cross-section just above threshold that is much steeper 
than phase-space expectations is discerned, which was not seen by
Belle due to limitations of the ISR method.
BESIII's measured cross section lineshape is different from Belle's, 
disfavoring a resonance like Y(4660) in the $\Lambda_c^+ \bar{\Lambda}_c^-$ 
channel. The BESIII results have driven discussions in the 
theoretical literature~\cite{Dai:2017fwx}.

The relatively larger samples at $\sqrt{s} = $ 4.5745 and 4.5995 GeV
enabled studies of the polar angular distribution of $\Lambda_c$ in the
$e^+e^-$ center-of-mass system. The shape function $f(\theta) \propto
(1 + \alpha_{\Lambda_c}\cos^2\theta)$ is fitted to the combined data contaning
the yields of $\Lambda_c^+$ and $\bar{\Lambda}_c^-$ for all ten decay modes
as shown in Fig.~\ref{lambdac2}(b). 
The ratio between the electric and magnetic form factors $|G_E/G_M|$ 
can be extracted using
$|G_E/G_M|^2(1-\beta^2) = (1 - \alpha_{\Lambda_c})/ (1 + \alpha_{\Lambda_c})$.
From these distributions, the ratios $|G_E/G_M|$ of $\Lambda_c^+$ 
have been extracted for the first time: they are
$1.14 \pm 0.14~(\rm stat.) \pm 0.07~(\rm sys.)$ and 
$1.23 \pm 0.05~(\rm stat.) \pm 0.03~(\rm sys.)$ 
at $\sqrt{s} = $ 4.5745 and 4.5995 GeV, respectively.    

At a future STCF with energy up to 7 GeV, it is natural to extend
the studies to other charmed baryons including $\Sigma_c$, $\Xi_c$,
$\Omega_c$ and their excited states.

\section{Summary}
For a long time the overall uncertainties on two of the fundamental
quantities $\alpha(M_Z)$ and $a_{\mu}$ had been dominated by the 
poor precision (15-20\% accuracy) of the $R$ values at the energies 
between $1-5$ GeV. BESII reduced the uncertainty of $R$ values in 
$2-5$ GeV to a level of $6-7\%$, which helped a lot in the 
Standard Model fitting. Recently BESIII has reached a precision 
better than 3\% in the continuum region. 
Nonetheless, with the improvement of measurement precision 
in other energy regions as well, especially in the region 
below 2 GeV, once again the uncertainty of $R$ values in $2-5$ GeV
becomes a limiting factor. 
In view of the importance of accurate values of $\alpha(M_Z)$ and $a_{\mu}$
for the precision test of the SM, new measurements of $R$ with a 
precision towards $(1\sim3)\%$ are strongly encouraged and 
called for at future STCF. 
Such a measurement is also important for the evaluation of the
strong coupling constant $\alpha_s$ and the charm quark mass.

It will be a very difficult task to measure $R$ to a precision 
of a percent in the energy region of $2-5$ GeV.  
A great effort has to be made to reduce the uncertainty in 
the detection efficiency for the hadronic events and
luminosity measurement, and beam associated background 
and initial state radiative correction also have to be 
examined carefully.

Hadron production and hadron structure are yet to be further studied,
preferably with high-quality and large-statistics data at a future STCF.
Some important topics, like form factor measurement, production threshold
behavior study, fragmentation function investigation, etc.,
are essential to understand the strong interaction ultimately.


\begin{thebibliography}{99}

\bibitem{STCF}
Q. Luo and D. Xu,
Progress on Preliminary Conceptual Study of HIEPA, a Super Tau-Charm Factory in China.
9th International Particle Accelerator Conference (IPAC 2018),
Vancouver, Canada, April 29-May 4, 2018.

\bibitem{SCTF}
A. Y. Barnyakov, 
The project of the Super Charm-Tau Factory in Novosibirsk.
J.\ Phys.\ Conf.\ Ser.\  2020; {\bf 1561}: 012004.

\bibitem{EW1} S. L. Glashow, Nucl.. Phys. 22 (1961) 579
\bibitem{EW2} S. Weinberg, Phys. Rev. Lett. 19 (1967) 124
\bibitem{alphaR} N. Cabibbo and R. Catto, Phys. Rev. 124 (1961)1577 
\bibitem{HVP2018} A. Keshavarzi, D. Nomura and T. Teubner, Phys. Rev. D97, 114025 (2018)
\bibitem{mucern1} J. Bailey et al., Phys. Lett. B 68 (1977), 
\bibitem{mucern2} F.J.M. and E. Picasso, Ann. Rev. Nucl. Sci. 29 (1979) 243
\bibitem{mucern3} F.J.M. Farley, Z. Phys. C 56 (1992) S88 
\bibitem{E821} G.W. Bennett et al. (Muon g-2 Collaboration), Phys.Rev.Lett. 92 (2004) 161802, Phys. Rev. D 73, 072003 (2006).
\bibitem{amutheo1} D.J. Broadhurst, et al., Phys. Lett. B 298 (1993) 445
\bibitem{amutheo2} T. Kinoshita Phys. Rev. D 47 (1993) 5013
\bibitem{amutest1} T. Kinoshita and W.J. Marciano, in 'Quantum Electrodynamics', ed. T. Kinoshita, World Scientific, Singapore, 1990, pp. 419-478
\bibitem{amutest2} P. Mery, et al., Z. Phys. C 46 (1990) 229
\bibitem{amutest3} W. Bernreuther, Z. Phys. C 56 (1992) S97
\bibitem{amutest4} D. H. Brown et al., Phys. Rev. D 54 (1996) 3237, A. Czarnecki, et al., 
'Electroweak corrections to the muon anomalous magnetic moment', hep-ph 9512369
\bibitem{amuSM2020} T.~Aoyama, N.~Asmussen, M.~Benayoun {\it et al.}, \href{https://doi.org/10.1016/j.physrep.2020.07.006}{Phys. Rep. {\bf 887}, 1 (2020)}.
\bibitem{amu2021FNAL} B.~Abi {\it et al.} (Muon $g-2$ Collaboration), \href{https://doi.org/10.1103/PhysRevLett.126.141801}{Phys. Rev. Lett. {\bf 126}, 141801 (2021)}.

\bibitem{RQCD} S. G. Gorishny et. al., Phys. Lett. B259 (1991) 144
   R. Marshall, Z. Phys. C 43 (1988) 595
\bibitem{PDG21} P.A. Zyla et al. (Particle Data Group), Prog. Theor. Exp. Phys. 2020, 083C01 (2020) and 2021 update
\bibitem{Rref1} $\mu\pi$ group: F. Ceradini et al., Phys. Lett. B 47 (1973) 80;
   $\gamma\gamma$ group: C. Bacci et al., ibid. B 44 (1973) 533;
   Boson group: B. Batoli et al., Phys. Rev. D 6 (1972) 2374;
   BCF group: M. Bernardini et al., Phys. Lett. B 51 (1974) 200
\bibitem{Rref2} G. Cosme et al., Phys. Lett. B 40 (1972) 685 and references herein 
\bibitem{Rref3} M. Kurdadze et al., Phys. Lett. B 42 (1972) 515
\bibitem{Rref4} Litke et al., Phys. Rev. Lett. 30 (1973) 1189 
   Phys. Rev. Lett. 30 (1973) 1349
\bibitem{bes2r} BES Collaboration, J. Z. Bai et al., Phys. Rev. Lett. 84, 594 (2000);
BES Collaboration, J. Z. Bai et al., Phys. Rev. Lett. 88, 101802 (2002)
\bibitem{DM1} A. E. Blinov et al., Z.Phys. C70 (1996) 31
\bibitem{CLEO3} CLEO Collaboration, D. Besson et al., Phys. Rev. D76 (2007) 072008
\bibitem{KEDR2016} V. V. Anashin {\it et al.} (KEDR Collaboration), \href{https://doi.org/10.1016/j.physletb.2015.12.059}{Phys. Lett. B {\bf 753}, 533 (2016)}.
\bibitem{KEDR2017} V. V. Anashin {\it et al.} (KEDR Collaboration), \href{https://doi.org/10.1016/j.physletb.2017.04.073}{Phys. Lett. B {\bf 770}, 174 (2017)}.
\bibitem{KEDR2019} V. V. Anashin {\it et al.} (KEDR Collaboration), \href{https://doi.org/10.1016/j.physletb.2018.11.012}{Phys. Lett. B {\bf 788}, 42 (2019)}.
\bibitem{BESIII2022} M. Ablikim {\it et al.} (BESIII Collaboration), \href{https://doi.org/10.1103/PhysRevLett.128.062004}{Phys. Rev. Lett. {\bf 128}, 062004 (2022)}.
\bibitem{DASP} R. Brandelik et al. DASP Collaboration, Phys. Lett. B 76 (1978) 361
\bibitem{MarkI} J. L. Siegrist et al. Mark I Collaboration, Phys. Rev. D 26 (1982) 969 
\bibitem{PLUTO} J. Burmeister et al. PLUTO Collaboration, Phys. Lett. B 66 (1977) 395
\bibitem{respara} M. Ablikim et al., BES Collaboration, Phys. Lett. B660, 315 (2008)
\bibitem{CLEOc13} D. Cronin-Hennessy et al., CLEO Collaboration, Phys. Rev. D80, 072001 (2009)
\bibitem{BelleISR} G. Pakhlova, Belle Collaboration, International Workshop on $e^+e^-$ from Phi to Psi, 13-16 October, 2009, Beijing, China
\bibitem{BES09} M. Ablikim et al., BES Collaboration, Phys. Lett. B677 (2009) 239
\bibitem{alphas2020} Diogo Boito and Vicent Mateu, \href{https://doi.org/10.1007/JHEP03(2020)094}{JHEP 03, 094 (2020)}.
\bibitem{cmass1} J.H.K\"{u}hn and M.Steinhauser, Nucl. Phys. B619, 588 (2001)
\bibitem{cmass2} J.Pe\~{n}arrocha and K.Schilcher, Phys. Lett B515, 291 (2001)
\bibitem{MLLA} V.A. Khoze and W. Ochs, Int. J. Mod. Phys. A 12, 2949 (1997).
\bibitem{LPHD} Ya.I. Azimov, Yu.L. Dokshitzer, V.A. Khoze, and S.I. Troyan, Phys. Lett. B165, 147 (1985); Z. Phys. C 27, 65 (1985)
\bibitem{yanwb} J.Z. Bai, BES Collaboration, Phys. Rev. D 69, 072002 (2004)
\bibitem{lihh} M. Ablikim et al., BES Collaboration, Phys. Lett. B630 (2005) 14
\bibitem{dubnicka} A. Z. Dubnickova, S. Dubnicka and M.P. Rekalo, Nuovo Cim. A 109 (1996) 241.

\bibitem{strangeonium} D.B. Lichtenberg and J.G. Wills, Phys. Rev. Lett. {\bf 35} 1055 (1975).
\bibitem{Y2175} Dian-Yong Chen {\it et al.}, Eur. Phys. J. C {\bf 72}, 2008 (2012).
\bibitem{Y2175BABAR11} B.~Aubert {\it et al.} ({\it BABAR} Collaboration),  \href{https://journals.aps.org/prd/abstract/10.1103/PhysRevD.74.091103}{{\color {black}Phys.\ Rev.\ D\ {\bf 74}, 091103(R) (2006);}} \href{https://journals.aps.org/prd/abstract/10.1103/PhysRevD.76.012008}{{\color{black}{\bf 76}, 012008 (2007).}} 
\bibitem{Y2175BABAR12} J.~P.~Lees {\it et al.} ({\it BABAR} Collaboration), \href{https://journals.aps.org/prd/abstract/10.1103/PhysRevD.86.012008}{{\color{black}Phys.\ Rev.\ D\ {\bf 86}, 012008 (2012).}}
\bibitem{Y2175BELLE} C.~P.~Shen {\it et al.} (Belle Collaboration),  \href{https://journals.aps.org/prd/abstract/10.1103/PhysRevD.80.031101}{{\color {black}Phys.\ Rev.\ D\ {\bf 80}, 031101(R) (2009)}}.
\bibitem{Y2175BESII} M.~Ablikim {\it et al.} (BES Collaboration), \href{https://journals.aps.org/prl/abstract/10.1103/PhysRevLett.100.102003}{{\color{black} Phys.\ Rev.\ Lett.\ {\bf 100}, 102003 (2008).}}
\bibitem{Y2175BESIII} M.~Ablikim {\it et al.} (BESIII Collaboration), \href{https://journals.aps.org/prd/abstract/10.1103/PhysRevD.91.052017}{{\color{black}Phys. \ Rev. \ D \ {\bf 91}, 052017 (2015).}}
\bibitem{Y2175BESIII2019} M.~Ablikim {\it et al.} (BESIII Collaboration), \href{https://journals.aps.org/prd/abstract/10.1103/PhysRevD.99.012014}{{\color{black}Phys. \ Rev. \ D \ {\bf 99}, 012014 (2019).}}
\bibitem{Y2175SSbar2} T.~Barnes, N.~Black, and P.~R.~Page, \href{https://journals.aps.org/prd/abstract/10.1103/PhysRevD.68.054014}{{\color{black}Phys. Rev. D {\bf 68}, 054014 (2003).}}
\bibitem{Y2175ss2D} G.~J.~Ding and M.~L.~Yan, \href{https://www.sciencedirect.com/science/article/pii/S0370269307012245?via\%3Dihub}{{\color{black}Phys.\ Lett.\ B \ {\bf 657}, 49 (2007)}}; Q.~Li, L.~C.~ Gui, M.~S.~ Liu, Q.~F.~ Lv, and X.~H.~ Zhong, \href{https://arxiv.org/abs/2004.05786}{{\color{black} arXiv: 2004.05786}}.
\bibitem{Y2175hybrid} G.~J.~Ding and M.~L.~Yan, \href{https://www.sciencedirect.com/science/article/pii/S0370269307006107?via\%3Dihub}{{\color{black}Phys.\ Lett.\ B \ {\bf 650}, 390 (2007).}}
\bibitem{Y2175hybrid3} P.~R.~Page, E.~S.~Swanson, and A.~P.~Szczepaniak, \href{https://journals.aps.org/prd/pdf/10.1103/PhysRevD.59.034016}{{\color{black}Phys. Rev. D {\bf 59}, 034016 (1999).}}
\bibitem{Y2175tetraquark1} Z.~G.~Wang, \href{https://www.sciencedirect.com/science/article/pii/S0375947407004629?via\%3Dihub}{{\color {black}Nucl.\ Phys.\ {\bf A 791}, 106 (2007).}}
\bibitem{Y2175tetraquark2} H.~X.~Chen, X.~Liu, A.~Hosaka, and S.~L.~Zhu, \href{https://journals.aps.org/prd/abstract/10.1103/PhysRevD.78.034012}{{\color{black}Phys.\ Rev.\ D \ {\bf 78}, 034012 (2008).}}
\bibitem{Y2175tetraquark3} N.~V.~Drenska, R.~Faccini, and A.~D.~Polosa, \href{https://www.sciencedirect.com/search?pub=Physics\%20Letters\%20B&volume=669&page=160&show=25&sortBy=relevance&origin=jrnl_issue&zone=search&cid=271623}{{\color{black}Phys. \ Lett. \ B {\bf 669}, 160 (2008).}}
\bibitem{Y2175tetraquark4} H.~W.~Ke and X.~Q.~Li, \href{https://journals.aps.org/prd/pdf/10.1103/PhysRevD.99.036014}{ {\color {black}Phys.\ Rev.\ D \ {\bf 99}, 036014 (2019).}}
\bibitem{Y2175tetraquark5} S.~S.~Agaev, K.~Azizi, and H.~Sundu, \href{https://journals.aps.org/prd/abstract/10.1103/PhysRevD.101.074012}{{\color {black} Phys. \ Rev. \ D \ {\bf 101}, 074012 (2020)}}.
\bibitem{Y2175lambda} E.~Klempt and A.~Zaitsev, \href{https://www.sciencedirect.com/science/article/pii/S0370157307003560?via\%3Dihub}{{\color{black}Phys.\ Rep.\ {\bf 454}, 1 (2007).}}
\bibitem{Y2175lambda1} C.~F.~Qiao, \href{https://www.sciencedirect.com/science/article/pii/S0370269306007696?via\%3Dihub}{{\color{black}Phys. \ Lett. \ B {\bf 639}, 263 (2006).}}
\bibitem{Y2175lambda2} Y.~B.~Dong {\it et al.,} \href{https://journals.aps.org/prd/abstract/10.1103/PhysRevD.96.074027}{{\color{black}Phys.\ Rev.\ D \ {\bf 96}, 074027 (2017).}}
\bibitem{Y2175lambda3} Y.~L.~Yang, D.~Y.~Chen, and Z.~Lu,\href{https://journals.aps.org/prd/abstract/10.1103/PhysRevD.100.073007}{{\color {black}Phys.\ Rev.\ D \ {\bf 100}, 073007 (2019)}}.
\bibitem{SWaveThreshold} S.~L.~Zhu, \href{https://www.worldscientific.com/doi/abs/10.1142/S0218301308009446}{{\color{black}Int. J. Mod. Phys. E {\bf 17}, 283 (2008).}}
\bibitem{X2170} A.~M.~Torres, K.~P.~Khemchandani, L.~S.~Geng, M.~Napsuciale, and E.~Oset, \href{https://journals.aps.org/prd/abstract/10.1103/PhysRevD.78.074031}{{\color{black}Phys.\ Rev.\ D\ {\bf 78}, 074031 (2008).}}
\bibitem{KK} M. Ablikim {\it et al.} (BESIII Collaboration), Phys. Rev. D {\bf 99}, 032001 (2019).
\bibitem{KSKL} M. Ablikim {\it et al.} (BESIII Collaboration), Phys. Rev. D {\bf 104}, 092014 (2021).
\bibitem{KKpi0pi0} M. Ablikim {\it et al.} (BESIII Collaboration), Phys. Rev. Lett. {\bf 124}, 112001 (2020).
\bibitem{KKKK} M. Ablikim {\it et al.} (BESIII Collaboration), Phys. Rev. D {\bf 100}, 032009 (2019).
\bibitem{phieta} M. Ablikim {\it et al.} (BESIII Collaboration), Phys. Rev. D {\bf 104}, 032007 (2021).
\bibitem{phietap} M. Ablikim {\it et al.} (BESIII Collaboration), Phys. Rev. D {\bf 102}, 012008 (2020).
\bibitem{omegaeta} M. Ablikim {\it et al.} (BESIII Collaboration), Phys. Lett. B {\bf 813}, 136059 (2021).

\bibitem{martin} Quarks and leptons: an introductory course in modern particle physics, by Francis Halzen and Alan D. Martin

\bibitem{ppbarBaBar1}
Lees JP, Poireau V and Tisserand V {\it et al.} [BaBar Collaboration].
Study of $e^+e^- \to p \bar{p}$ via initial-state radiation at BABAR.
Phys. Rev. D 2013; {\bf 87}: 092005.

\bibitem{ppbarBaBar2}
Lees JP, Poireau V and Tisserand V {\it et al.} [BaBar Collaboration].
Measurement of the $e^+e^- \to p\bar{p}$ cross section in the energy range from 3.0 to 6.5 GeV.
Phys. Rev. D 2013; {\bf 88}: 072009.

\bibitem{lear} Bardin G, Burgun G, Calabrese R, Capon G, Carlin R (1994) {\it et al.}{\it Nucl. Phys.} B {\bf 411} 3-32
\bibitem{Feni} Antonelli A {\it et al} 1998 {\it Nucl. Phys.} B {\bf 517} 3 
\bibitem{crosssection} Zichichi A, Berman S M, Cabibbo N and Gatto R 1962 {\it Nuovo Cimento} {\bf 24} 170 
\bibitem{Tzara} Tzara C 1970 {\it Nucl. Phys.} B {\bf 18} 216-252

\bibitem{ppbar2015}
Ablikim M, Achasov MN and Ai XC{\it et al.} [BESIII Collaboration].
Measurement of the proton form factor by studying $e^{+} e^{-}\rightarrow p\bar{p}$.
Phys. Rev. D 2015; {\bf 91}: 112004.

\bibitem{ppbar2020}
Ablikim M, Achasov MN and Adlarson P {\it et al.} [BESIII Collaboration].
Measurement of proton electromagnetic form factors in $e^+e^- \to p\bar{p}$ in the energy region 2.00 - 3.08 GeV.
Phys. Rev. Lett. 2020; {\bf 124}: 042001.

\bibitem{ppbaruntag}
Ablikim M, Achasov MN and Adlarson P {\it et al.} [BESIII Collaboration].
Study of the process $e^+ e^- \to p \bar p$ via initial state radiation at BESIII.
Phys. Rev. D 2019; {\bf 99}: 092002.

\bibitem{ppbartag}
Ablikim M, Achasov MN and Adlarson P {\it et al.} [BESIII Collaboration].
Measurement of proton electromagnetic form factors in the time-like region using initial state radiation at BESIII.
Phys. Lett. B 2021; {\bf 817}: 136328.

\bibitem{CMD3}
Akhmetshin RR, Amirkhanov AN and Anisenkov AV {\it et al.} [CMD-3 Collaboration]. 
Observation of a fine structure in $e^+ e^- \to$ hadrons production at the nucleon-antinucleon threshold.
Phys. Lett. B 2019; {\bf 794}: 64-68.

\bibitem{unitary1}
Adamuscin C, Dubnicka S and Dubnickova AZ {\it et al.}
A unitary and analytic model of nucleon EM structure, the puzzle of JLab proton polarization data and new insight into the proton charge distribution.
Prog. in Parti. and Nucl. Phys. 2005; {\bf 55}: 228-241.

\bibitem{unitary2}
Dubnickova AZ and Dubnicka S.
Proton em form factors data are in disagreement with new $\sigma_{tot}(e^+e^- \to p\bar p)$ measurements.
arXiv:2010.15872 [hep-ph] (2020).

\bibitem{Dalkarov}
Dalkarov OD, Khakhulin PA and Voronin AY.
On the electromagnetic form factors of hadrons in the time-like region near threshold.
Nucl. Phys. A 2010; {\bf 833}: 104-118.

\bibitem{PeriodicFF} 
Bianconi A and Tomasi-Gustafsson E.
Periodic interference structures in the timelike proton form factor.
Phys. Rev. Lett. 2015; {\bf 114}: 232301.

\bibitem{nnbar}
Ablikim M, Achasov MN and Adlarson P {\it et al.} [BESIII Collaboration].
Oscillating features in the electromagnetic structure of the neutron.
Nat. Phys. 2021; {\bf 17}: 1200.

\bibitem{SND2014}
Achasov MN, Barnyakov AY and Beloborodov KI {\it et al.} [SND Collaboration]. 
Study of the process $e^+e^-\to n\bar{n}$ at the VEPP-2000 $e^+e^-$ collider with the SND detector.
Phys. Rev. D 2014; {\bf 90}: 112007.

\bibitem{SNDnn}
Druzhinin VP and Serednyakov SI.
Measurement of the $e^+e^- \to n \bar n$ cross section with the SND detector at the VEPP-2000 collider.
EPJ Web Conf. 2019; {\bf 212}: 07007.

\bibitem{DM2BB}
Bisello D, Busetto G and Castro A {\it et al.} [DM2 Collaboration].
Baryon pairs production in $e^+e^-$ annihilation at $\sqrt(s)=2.4$ GeV.
Z. Phys. C 1990; {\bf 48}: 23-28.

\bibitem{LamSigBaBar2007}
Aubert B, Bona M and Boutigny D {\it et al.} [BABAR Collaboration].
Study of $e^{+} e^{-} \to \Lambda \bar{\Lambda}$, $\Lambda \bar{\Sigma}^0$, $\Sigma^0 \bar{\Sigma}^0$ using initial state radiation with BABAR.
Phys. Rev. D 2007; {\bf 76}: 092006.

\bibitem{Lambdapair} Ablikim, M.; Achasov, M. N.; Ahmed, S.; et al.
Observation of a cross-section enhancement near mass threshold in 
$e^{+}e^{-}\rightarrow\Lambda\bar{\Lambda}$. 
\href{http://dx.doi.org/10.1103/PhysRevD.97.032013}{\emph{Phys.\ Rev.\ D} {\bf 2018}, \emph{97}, 032013.}

\bibitem{TheoRes}
El-Bennich B, Lacombe M and Loiseau B {\it et al.} 
Paris $N\overline{N}$ potential constrained by recent antiprotonic-atom data and $\overline{n}p$ total cross sections.
Phys. Rev. C 2009; {\bf 79}: 054001.

\bibitem{Xiao:2019qhl}
Xiao LY, Weng XZ and Zhong XH {\it et al.} 
A possible explanation of the threshold enhancement in the process $e^+e^-\rightarrow \Lambda\bar{\Lambda}$.
Chin. Phys. C 2019; {\bf 43}: 113105.

\bibitem{Resumfct1}
Baldini R, Pacetti S and Zallo A {\it et al.} 
Unexpected features of $e^+e^- \rightarrow$p$ \overline{{p}}$ and $e^+e^- \rightarrow \Lambda \overline{\Lambda}$ cross-sections near threshold.
Eur. Phys. J. A 2009; {\bf 39}: 315-321.

\bibitem{Resumfct2}
Baldini Ferroli R, Pacetti S and Zallo A.
No Sommerfeld resummation factor in $e^+e^- -> p \bar{p}$~?
Eur. Phys. J. A 2012; {\bf 48}: 33.

\bibitem{FSinter1}
Zou BS and Chiang HC.
One-pion-exchange final-state interaction and the $p\overline{p}$ near threshold enhancement in $J/\psi\to\gamma p\overline{p}$ decays.
Phys. Rev. D 2004; {\bf 69}: 034004.

\bibitem{FSinter2}
Haidenbauer J, Hammer HW and Mei\ss{}ner UG {\it et al.} 
On the strong energy dependence of the $e^+ e^- \leftrightarrow p \bar{p}$ amplitude near threshold.
Phys. Lett. B 2006; {\bf 643}: 29-32.

\bibitem{FSinter3}
Haidenbauera J and Mei\ss{}ner UG.
The electromagnetic form factors of the $\Lambda$ in the timelike region.
Phys. Lett. B 2016; {\bf 761}: 456-461.

\bibitem{Sommerfeld} A. D. Sakharov, Zh. Eksp. Teor. Fiz. {\bf 18}, 631 (1948) 
[Sov. Phys. Usp. 34, 375 (1991)]; A. Sommerfeld, Atombau und Spektralliniem 
(Vieweg, Braunschweig, 1944), Vol. 2, p.130; 
J. Schwinger, Particles, Sources, and Fields, Vol. III, p. 80.

\bibitem{Rinaldo} R.~Baldini, S.~Pacetti, A.~Zallo and A.~Zichichi,
  Eur.\ Phys.\ J.\ A {\bf 39} (2009) 315
  [arXiv:0711.1725 [hep-ph]].

\bibitem{SigmaPLB2021} Ablikim, M.; Achasov, M.N.; Adlarson, P.; et al.
Measurements of~$\Sigma^{+}$ and $\Sigma^{-}$ timelike electromagnetic 
form factors for center-of-mass energies from 2.3864 to 3.0200 GeV. 
\href{http://dx.doi.org/10.1016/j.physletb.2021.136110}{\emph{Phys.\ Lett.\ B} {\bf 2021}, \emph{814}, 136110.}

\bibitem{Sigma0PLB2022} Ablikim, M.; Achasov, M.N.; Adlarson, P.; et al.
Measurement of the $e^+e^-\to\Sigma^{0}\bar{\Sigma}^{0}$ cross sections 
at center-of-mass energies from $2.3864$ to $3.0200$ GeV. 
arXiv:2110.04510v1.

\bibitem{chargedXi}
Ablikim M, Achasov MN and Adlarson P {\it et al.} [BESIII Collaboration].
Measurement of cross section for $e^+e^-\to\Xi^-\bar{\Xi}^+$ near threshold at BESIII.
Phys. Rev. D 2021; {\bf 103}: 012005.

\bibitem{neutralXi}
Ablikim M, Achasov MN and Adlarson P {\it et al.} [BESIII Collaboration].
Measurement of cross section for $e^{+}e^{-}\rightarrow\Xi^{0}\bar{\Xi}^{0}$ near threshold.
Phys. Lett. B 2021; {\bf 820}: 136557.

\bibitem{NSR2021} Huang,~G.S.; Baldini Ferroli,~R. 
Probing the internal structure of baryons.
\href{http://dx.doi.org/10.1093/nsr/nwab187}{{\em Natl. Sci. Rev.} {\bf 2021}, {\em 8}, nwab187.}

\bibitem{LambdacBelle}
Pakhlova G, Adachi I and Aihara H {\it et al.} [Belle Collaboration].
Observation of a Near-Threshold Enhancement in the ${e}^{+}{e}^{\ensuremath{-}}\ensuremath{\rightarrow}{\ensuremath{\Lambda}}_{c}^{+}{\ensuremath{\Lambda}}_{c}^{\ensuremath{-}}$ Cross Section Using Initial-State Radiation.
Phys. Rev. Lett. 2008; {\bf 101}: 172001.

\bibitem{Lambdacpair}
Ablikim M, Achasov MN and Ahmed S {\it et al.} [BESIII Collaboration].
Precision measurement of the $e^{+}e^{-}~\rightarrow~\Lambda_{c}^{+} \bar{\Lambda}_{c}^{-}$ cross section near threshold.
Phys. Rev. Lett. 2018; {\bf 120}: 132001.

\bibitem{dm2} D. Bisello \textit{et al.}, Z. Phys. C 48 (1990) 23.
\bibitem{czyz} H. Czyz, A. Grzelinska and J.H. K\"{u}hn, Phys. Rev. D \textbf{75} (2007) 074026.
\bibitem{korner} J. G. K\"{o}rner and M. Kuroda, Phys. Rev. D. 16 (1977) 2165.
\bibitem{bartos} E. Bartos, A. Z. Dubnickova and S. Dubnicka, Nucl. Phys. B (Proc. Suppl.) 219-220 (2011) 166.
\bibitem{dubnickova} S. Dubnicka and A. Z. Dubnickova, Acta Phys. Slov. 60 (2010) 1.
\bibitem{dubnickahep} S. Dubnicka \textit{et al.}, talk at the EPS HEP Conference in Stockholm, Sweden (2013).

\bibitem{LambdaPRL2019} Ablikim, M.; Achasov, M.N.; Adlarson, P.; et al.
Complete measurement of the $\Lambda$ electromagnetic form factors. 
\href{http://dx.doi.org/10.1103/PhysRevLett.123.122003}{\emph{Phys.\ Rev.\ Lett.}  {\bf 2019}, \emph{123}, 122003.}

\bibitem{LambdaPolar}
Ablikim M, Achasov MN and Ahmed S {\it et al.} [BESIII Collaboration].
Polarization and Entanglement in Baryon-Antibaryon Pair Production in Electron-Positron Annihilation.
Nature Physics 2019; {\bf 15}: 631-634.

\bibitem{alphaLambda}
Ireland DG, D\"oring M and Glazier DI {\it et al.} 
Kaon Photoproduction and the $\Lambda$ Decay Parameter $\alpha_-$.
Phys. Rev. Lett. 2019; {\bf 123}: 182301.

\bibitem{Dai:2017fwx}
Dai LY, Haidenbauer J and Mei\ss{}ner UG.
Re-examining the $X(4630)$ resonance in the reaction $e^+e^-\rightarrow \Lambda^+_c\bar\Lambda^-_c$.
Phys. Rev. D 2017; {\bf 96}: 116001.

\end{thebibliography}
\end{document}